\documentclass[aps,prd,preprint,nofootinbib,floatfix,amsmath,amssymb,groupedaddress]{revtex4-1}

\usepackage{hyperref}

\usepackage[utf8]{inputenc}
\usepackage[utf8]{inputenc}
\usepackage[T1]{fontenc}

\usepackage[usenames,dvipsnames]{xcolor}

\usepackage{pgf}

\usepackage{booktabs}

\usepackage{slashed}

\usepackage{rotating}

\usepackage{colortbl}

\usepackage{url}

\usepackage[retainorgcmds]{IEEEtrantools}

\usepackage{amsmath, amsthm, amsfonts,amssymb}

\usepackage{graphicx}

\usepackage{appendix}

\usepackage{braket}

\usepackage{feynmp}

\usepackage{ifpdf}
\ifpdf
\fi

\makeatletter
\def\endfmffile{%
  \fmfcmd{\p@rcent\space the end.^^J%
          end.^^J%
          endinput;}%
  \if@fmfio
    \immediate\closeout\@outfmf
  \fi
  \IfFileExists{\thefmffile.mp}{\immediate\write18{mpost \thefmffile}}{}
  \let\thefmffile\relax
}
\makeatother

\usepackage{tikz}
\usetikzlibrary{arrows,shapes}
\usetikzlibrary{trees}
\usetikzlibrary{matrix,arrows} 				
\usetikzlibrary{positioning}				
\usetikzlibrary{calc,through}				
\usetikzlibrary{decorations.pathreplacing}  
\usepackage{pgffor}							

\usetikzlibrary{decorations.pathmorphing}	
\usetikzlibrary{decorations.markings}
\tikzset{
    vector/.style={decorate, decoration={snake}, draw},
	provector/.style={decorate, decoration={snake,amplitude=2.5pt}, draw},
	antivector/.style={decorate, decoration={snake,amplitude=-2.5pt}, draw},
    fermion/.style={draw=black, postaction={decorate},
        decoration={markings,mark=at position .55 with {\arrow[draw=black]{>}}}},
    fermionbar/.style={draw=black, postaction={decorate},
        decoration={markings,mark=at position .55 with {\arrow[draw=black]{<}}}},
    fermionnoarrow/.style={draw=black},
    gluon/.style={decorate, draw=black,
        decoration={coil,amplitude=4pt, segment length=5pt}},
    scalar/.style={dashed,draw=black, postaction={decorate},
        decoration={markings,mark=at position .55 with {\arrow[draw=black]{>}}}},
    scalarbar/.style={dashed,draw=black, postaction={decorate},
        decoration={markings,mark=at position .55 with {\arrow[draw=black]{<}}}},
    scalarnoarrow/.style={dashed,draw=black},
    electron/.style={draw=black, postaction={decorate},
        decoration={markings,mark=at position .55 with {\arrow[draw=black]{>}}}},
	bigvector/.style={decorate, decoration={snake,amplitude=4pt}, draw},
}

\tikzstyle{block} = [draw, rectangle, 
    minimum height=3em, minimum width=6em]

\begin{document}


\title{$ \mathcal{N}= 1 $ super Feynman rules for any superspin: Noncanonical SUSY.}


\author{Enrique Jim\'enez.}
\email[]{ejimenez@fisica.unam.mx}
\affiliation{Instituto de F\'isica, Universidad Nacional Aut\'onoma de M\'exico, \\ \normalsize
 Apartado  Postal 20-364, 01000, M\'exico D.F., M\'exico}


\date{\today}

\begin{abstract}
Super Feynman rules for any superspin are given for massive $ \mathcal{N}=1 $ supersymmetric theories, including
momentum superspace on-shell legs. This is done by extending, from space to superspace, Weinberg’s
perturbative approach to quantum field theory. Superfields work just as a device that allow one to write
super Poincar\'e-covariant superamplitudes for interacting theories, relying neither in path integral nor
canonical formulations. Explicit transformation laws for particle states under finite supersymmetric
transformations are offered.  $ \mathit{C}, \mathit{P}, \mathit{T}, $ and $ \mathcal{R} $ transformations are also worked out. A key feature of this formalism is that it does not require the introduction of auxiliary fields, and when introduced, their purpose
is just to render supersymmetric invariant the time-ordered products in the Dyson series. The formalism is
tested for the cubic scalar superpotential. It is found that when a superparticle is its own antisuperparticle
the lowest-order correction of time-ordered products, together with its covariant part, corresponds to the
Wess–Zumino model potential. 
\end{abstract}


\maketitle
		\pagebreak  
\section{Introduction.}
\label{sec:intro}
From the inception of superspace by Salam and Strahdee\cite{Salam:1974yz},  functional   and path integral methods have been  the preferred scheme to formulate field theory in superspace\cite{Salam:1974pp,Salam:1974jj, Grisaru:1979wc}.  These formalisms allow us to write correlation functions that perturbatively give super Feynman rules with off-shell legs, making it unclear how to replace them  by the corresponding momentum superspace on-shell legs. Perhaps, because  realistic supersymmetric theories would never be symmetries of the S-matrix \cite{Dimopoulos:1981zb}, this issue seems secondary. However, thinking  of supersymmetry as a  theoretical laboratory, the issue has its own importance. A purpose of this paper is to provide formulas for on-shell legs in order to construct superamplitudes $ S_{\mathcal{N}\mathcal{M}} $ for scattering processes of massive superparticle states (or particle superstates), where $ \mathcal{N}  $ and $ \mathcal{M}  $ label Fock states, extended  such that one superparticle carries momentum  $ \mathbf{p}$,  spin-projection $ \sigma $, and left or right fermionic 4-spinors $ s_{+} $  or   $ s_{-} $.  These superamplitudes are constructed extending  Weinberg's approach\cite{Weinberg:1964cn,Weinberg:1969di} from  fields to superfields, that is from  (momentum and configuration) space to superspace. What is done here  is to  express  the potential appearing in the Dyson operator series 
\begin{IEEEeqnarray}{rl}
            S  \, = \, \mathcal{T}\exp\left[  \, - \,i\int dt \, V(t)\right] 
    \label{DysonRelation}
\end{IEEEeqnarray}
as
\begin{IEEEeqnarray}{rl}
            V(t)  \, = \, \int d^{3}{x}\,d^{4}\vartheta \,\mathcal{V}\left(x, \vartheta \right),
    \label{PotentialAsIntegralOfTheSuperPotential}
\end{IEEEeqnarray}
where  $ \mathcal{V}\left(x, \vartheta \right) $ is a sum of free superfield products obtained as super momentum Fourier transforms of creation-annihilation superparticle operators. These creation-annihilation superparticle operators are used to write superparticle states that allow us to write $S_{\mathcal{N}\mathcal{M}} $ in terms of super Feynman rules, after the appropriate Wick`s pairings. 
As in the ordinary space approach~\cite{Weinberg:1964cn}, the  assumed conditions for the super S-matrix are: perturbativity, unitarity, Poincar\'e covariance and clustering, with the addition of supersymmetry covariance. All of these are satisfied (with an important qualification made below) by Eqs. \eqref{DysonRelation} and \eqref{PotentialAsIntegralOfTheSuperPotential}.

One advantage of Weinberg's approach is that it represents an alternative perturbative formulation for massive quantum field theories, independently of whether a corresponding  canonical and/or path-integral formulation can be established~\footnote{For a discussion on these matters see Chapter 7 in~\cite{Weinberg:1995mt}. }. 
At  present, a systematic formulation to obtain general  massive super  Feynman rules from  canonical and/or path-integral formulations is not only unknown~\cite{Gates:2001rn}, but also only a few low superspin massive free  Lagrangians have been constructed\cite{Buchbinder:2002gh,Buchbinder:2002tt,Gregoire:2004ic,Gates:2013tka} (propagating component free fields for general massive supersymmetric multiples have been recently presented in \cite{Zinoviev:2007js}). Thus, one of the main aims of this paper is to provide a set of general super Feynman rules for massive  arbitrary superspins, where the hypothetical  canonical/path integral formulations from  which the rules can be derived are lacking (if they exist at all).   Since another aspect of  Weinberg's approach is that it tells us  what to expect from any massive field theory when considered in the interaction picture,  we hope that this new formulation will  provide  guidance for studies on the broader task of finding if indeed a systematic canonical and/or path-integral  formulation is possible\cite{Gates:2001rn}. 
  
This extension maintains all the properties of Weinberg's approach, i.e. super Feynman rules can be built for any superspin in a straightforward manner and one can easily incorporate charge conjugation, parity, time reversal and $ \mathcal{R} $-symmetries. Furthermore, it also allows us to obtain economic and concise expressions.

A characteristic feature of supersymmetric theories\cite{Wess:1973kz} is  that, when the Lagrangian does not contain auxiliary fields, the potential becomes not only a function of the coupling constant  $g$, but also of its square $g^{2}$, relating one and the next order in perturbation theory (otherwise  'miraculous' cancellations could not occur). Thus it is difficult to see how a perturbative scheme can cope with this situation. As in the case of Lorentz invariance, in considering  $  \mathcal{V}\left(x^{\mu}, \vartheta \right)  $  as an  invariant density under supersymmetry transformations,
\begin{IEEEeqnarray}{rl}
            \mathsf{U}(\xi) \mathcal{V}\left(x^{\mu}, \vartheta \right)  \mathsf{U}(\xi)^{-1}  \, = \,  \mathcal{V}\left(x^{\mu}+\vartheta^{\intercal}\epsilon\gamma_{5}\gamma^{\mu}\xi, \vartheta +\xi \right) \ ,
    \label{DensitySuperSymmetricTransformation}
\end{IEEEeqnarray}
is not sufficient to render supersymmetric invariant the time-ordered products appearing in  Eq. \eqref{DysonRelation}; therefore, we must introduce noncovariant  terms of higher order in coupling constants. We show that these  noncovariant terms are always local in space, making the  definition of the covariant super S-matrix possible\cite{Weinberg:1969di}.  For this perturbative  formalism, this seems to be the origin of auxiliary fields.\\

We adopt the notation and conventions of ~\cite{Weinberg:1995mt,Weinberg:2000cr}, except for left and right 4-spinors, which we write as $ 2\vartheta_{\pm}  \, = \, \left( I\pm \gamma_{5} \right)\vartheta  $ instead of $ \vartheta_{_{L,R}} $. As for the methods employed we use the standard techniques of  the operators' formalism and calculus in superspace (see for example~\cite{Weinberg:2000cr,Buchbinder:1995uq}). We present notation and all our conventions in Appendix~\ref{Sec_ConventionsNotation}. Also, we conjugate under the integrals of the fermionic variables and explain this in Appendix \ref{Sec_BerezinIntegrals}.

The article is structured as follows: In  Section \ref{Sec_SuperStates} unitary representations of the super Poincar\'e group are constructed.  Section  \ref{Sec_CausaleSuperfields}  deals with causal superfields,  and meanwhile  Section \ref{Sec_TimeOrderedSuperpropagators}   is devoted to time-ordered products and superpropagators. In Section \ref{Sec_SuperFeynmanRules} super Feynman rules are presented.  Charge conjugation, parity, time-reversal and $ \mathcal{R} $  transformation  formulas are written in Section \ref{Sec_CPTR}. The details  of the cubic superpotential  for a scalar superfield  are worked out in Section \ref{Sec_ScalarSuperPotential}. Finally, our conclusions are presented in Section \ref{Sec_Conclusions}.

\section{Creation-annihilation superparticle operators.}\label{Sec_SuperStates}
$\mathcal{N}=1$ supersymmetric multiplets  have four particle states with angular momentum $ (j,j, j\pm \frac{1}{2}) $~\footnote{Except for the case $ j=0 $. We call superspin $ j $ to the  set $ \left\lbrace j,j, j\pm\frac{1}{2}\right\rbrace $.}. With this in mind, we embed these states into two superparticle states,  one with left 4-spinor $s_{+} $ and the other with right 4-spinor $s_{-} $, and their fermionic expansion coefficients represent the states of the supersymmetric multiplet. We show that  super Poincar\'e transformations are acting unitarily on these superstates,  with the additional feature that  finite supersymmetric transformations are also considered. To do so, instead of taking states with $ j+ \frac{1}{2} $ and $ j- \frac{1}{2} $ angular momentum, we take these states to be in the tensorial representation  $ j\otimes \frac{1}{2} $. That is, at the level of creation operators, we start with\footnote{All states are constructed from  $ a^{*}(\cdots)\ket{\text{VAC}} $, where $ \ket{\text{VAC}} $ is a  super Poincar\'e invariant vacuum. Here, we  denote the adjoint of an operator as $ * $. When the adjoint is accompanied by a transpose of some vector, we denote it by  $ \dagger $. }
   \begin{IEEEeqnarray}{l}
         a^{*}_{+}(\textbf{p},\sigma) , \quad a^{*}_{-}(\textbf{p},\sigma),\quad  l^{*}_{a}(\textbf{p},\sigma),\quad a \, = \,  +\tfrac{1}{2},-\tfrac{1}{2}\ , \quad 
     \label{OperatorState}
 \end{IEEEeqnarray}  
that satisfy the  (nonzero) (anti)commutators~\footnote{  $\left\lbrace   \,\,\right]  $ is defined to be an  anticommutation or commutation if $ \left[\,\, \right\rbrace $   is a commutation or  an anticommutation, respectively.  } 
 \begin{IEEEeqnarray}{rl}
                \left[a_{ \pm}(\textbf{p},\sigma),a^{*} _{\pm}(\textbf{p}',\sigma') \right\rbrace  & \, = \, \delta^{3}(\textbf{p}-\textbf{p}')\delta_{\sigma\sigma'}\ ,    \nonumber \\               
                \left\lbrace l_{ a}(\textbf{p},\sigma) ,l^{*}_{ b}(\textbf{p}',\sigma')\right]   & \, = \, \delta^{3}(\textbf{p}-\textbf{p}')\delta_{ab}\delta_{\sigma\sigma'} \ ,  \nonumber \\                                   
    \label{IndependentComponent_MassiveCase}
    \end{IEEEeqnarray}     
and under a Poincar\'e transformation behave as
\begin{IEEEeqnarray}{rl}
                                     \mathsf{U}(\Lambda,x)a^{*}_{\pm}(\textbf{p},\sigma)  \mathsf{U}(\Lambda,x)^{-1}&\, = \, e^{-i {\Lambda} p\cdot x} \sqrt{\tfrac{\left(\Lambda p\right) ^{0}}{p^{0}}}\sum_{\sigma'}U^{(j)}_{\sigma'\sigma}[W(\Lambda,\textbf{p})]a^{*}_{\pm}(\textbf{p}_{\Lambda},\sigma'), \nonumber \\   
                                                             \mathsf{U}(\Lambda,x)l^{*}_{a}(\textbf{p},\sigma)  \mathsf{U}(\Lambda,x)^{-1} &\, = \, e^{- i{\Lambda}p\cdot x}  \sqrt{\tfrac{\left(\Lambda p\right) ^{0}}{p^{0}}}\sum_{b,\sigma'}U^{(j)}_{\sigma'\sigma}[W(\Lambda,\textbf{p})]U^{({\frac{1}{2}})}_{ba}\left[ W(\Lambda,\textbf{p})\right] l^{*}_{b}(\textbf{p}_{\Lambda},\sigma'),  \nonumber \\
     \label{Massive_ComponentOperatorStatesTransformation}
\end{IEEEeqnarray}    
where $ U^{(j)} $ is the spin-$j$ rotation matrix and $  W(\Lambda,\textbf{p}) $ is the so-called Wigner rotation,
\begin{equation}
              W(\Lambda,\textbf{p})  \, = \, L(\Lambda p)^{-1}\Lambda L(p), \quad p =L(p)k\ ,
     \label{WignerTransformation}
 \end{equation}
 with $ k= \begin{pmatrix}
 0 & 0 & 0 & m
 \end{pmatrix}  $ as a standard vector and $ W(\Lambda,\textbf{p})$ isomorphic to the rotation group.  As a definition, fermionic (bosonic) creation-annihilation particle operators remain fermionic (bosonic) with respect to supernumbers. 
A very  important fact  is that when a Lorentz  transformation $ R $ is an element of the rotation group the following relation holds:
\begin{IEEEeqnarray}{l}
           \left[  D_{\pm}(R)\right] _{ab}  \, = \, U^{(\frac{1}{2})}_{ab}(R) \ ,  
    \label{RotationDiracAsUnitaryTransform}
\end{IEEEeqnarray}
where  $ D_{\pm} $ stands for  the Weyl  representations. We embed the operators $ l^{*}_{a} $ in a four component vector,
\begin{IEEEeqnarray}{l}
            {b(\textbf{p},\sigma)}   \, \equiv \,  D\left[ L\left( \textbf{p}\right) \right]\begin{pmatrix}
\, l(\mathbf{p},\sigma) \\ 
l(\mathbf{p},\sigma)
\end{pmatrix} \ , 
    \label{MojaranaSoperator}
\end{IEEEeqnarray}
with $ D[\Lambda]  \, = \, D_{+}(\Lambda)\oplus D_{-}(\Lambda)$, the Dirac representation. In view of  \eqref{Massive_ComponentOperatorStatesTransformation} and \eqref{RotationDiracAsUnitaryTransform}:
\begin{IEEEeqnarray}{l}
               \mathsf{U}(\Lambda,x)\overline{b}(\textbf{p},\sigma)  \mathsf{U}(\Lambda,x)^{-1}\, = \, e^{-i {\Lambda}p\cdot x}  \sqrt{\tfrac{\left(\Lambda p\right) ^{0}}{p^{0}}}\sum_{\sigma'}U^{(j)}_{\sigma'\sigma}[W(\Lambda,\textbf{p})] \overline{b}(\textbf{p}_{\Lambda},\sigma')D\left[ {\Lambda} \right]\ ,  \nonumber \\
     \label{MassiveFourComponentStatesPoincareTransformation}
\end{IEEEeqnarray} 
  where $ \bar{b} $ is  the Dirac adjoint $ b^{\dagger}\beta $. The  nonvanishing (anti)commutation relations of $\left(  {b} , \bar{b} \right)$ are 
\begin{IEEEeqnarray}{rl}         
        \left\lbrace     {b_{\alpha}(\textbf{p},\sigma)}, \overline{b}_{\beta}(\textbf{p}',\sigma')\right]   \, = \,\left[ I \, + \, \left(-i\slashed{p} \right)/m  \right]_{\alpha\beta} \,\delta(\textbf{p} -\textbf{p}')\delta_{\sigma\sigma'} \ .
     \label{Massive_MutationRel_bSpinor}
\end{IEEEeqnarray}
One can also show that
\begin{IEEEeqnarray}{rl}
       \left( -i\slashed{p}\right)   { b(\textbf{p},\sigma)} \, = \,  m\,{ b(\textbf{p},\sigma)}\ ,  
    \label{MassivePropertyOf_b_DiracType}
\end{IEEEeqnarray}
which is a reminder that, although we are  using a four dimensional vector with  $ 4(2j+1) $  spin projections, only $ 2(2j+1)  $ of them are independent.  

We define two types of creation superparticle (sparticle) operators:
\begin{IEEEeqnarray}{rl}
            a^{*}_{\pm}(\textbf{p},s_{\pm},\sigma)   &   \,\equiv \, a^{*}_{\pm}(\textbf{p},\sigma)   \, \pm \, \sqrt{{2m}} \,\overline{b}  (\textbf{p},\sigma) \,   {s}_{\pm} \, \pm \,  \, 2 m \,\delta^{2}(s_{\pm}) a^{*}_{\mp }(\textbf{p},\sigma) \ ,            
    \label{Massive_CreationAnnihilaitonSuperoperators_plusminus}
\end{IEEEeqnarray}
with their corresponding annihilation sparticle operators \begin{IEEEeqnarray}{rl}
            a_{\mp}(\textbf{p},s_{\mp},\sigma)   & \, \equiv \, \left( a^{*}_{\pm}(\textbf{p},(\epsilon\gamma_{5}\beta s^{*})_{\pm},\sigma)\right)^{*} \nonumber \\
&   \,= \,  a_{\pm}(\textbf{p},\sigma)  \, \pm\,\sqrt{{2}{m}} \, {s}_{\mp}^{\intercal}\epsilon\gamma_{5}   \, b(\textbf{p},\sigma)  \, \mp \,2m \,  \delta^{2}(s_{\mp}) a_{\mp }(\textbf{p},\sigma) \ .
    \label{Massive_AnnihilationSuperOperator}
\end{IEEEeqnarray}
Creation-annihilation sparticle operators  have  the Poincar\'e transformation property:
\begin{IEEEeqnarray}{rl}
               \mathsf{U}(\Lambda,x) a^{*}_{\pm}(\textbf{p},s_{\pm},\sigma)  \mathsf{U}(\Lambda,x)^{-1} &\, = \, e^{-i{\Lambda} p\cdot x}  \sqrt{\tfrac{\left(\Lambda p\right) ^{0}}{p^{0}}}\sum_{\sigma'}U^{(j)}_{\sigma'\sigma}[W(\Lambda,\textbf{p})]  a^{*}_{\pm}(\mathbf{p}_{\Lambda},D(\Lambda)s_{\pm},\sigma')\ ,  \nonumber \\
                \mathsf{U}(\Lambda,x) a_{\pm}(\textbf{p},s_{\pm},\sigma)  \mathsf{U}(\Lambda,x)^{-1} &\, = \, e^{+i{\Lambda} p\cdot x}  \sqrt{\tfrac{\left(\Lambda p\right) ^{0}}{p^{0}}}\sum_{\sigma'}U^{(j)*}_{\sigma'\sigma}[W(\Lambda,\textbf{p})]  a_{\pm}(\mathbf{p}_{\Lambda},D(\Lambda)s_{\pm},\sigma') \ , \nonumber \\
     \label{Massive_PoincareTrans_SuperCreationOperator}
\end{IEEEeqnarray} 
 and the (nonzero) anti(commutation) relations
\begin{IEEEeqnarray}{rl}
            \left[        a_{\mp}(\textbf{p},s_{\mp},\sigma)  ,   a^{*}_{\pm}(\textbf{p}',s_{\pm}',\sigma')\right\rbrace   &\, = \, \delta^{3}\left( \textbf{p}-\textbf{p}'\right) \delta_{\sigma\sigma'} \exp{  \left[  2 {s}^{\intercal}\epsilon\gamma_{5}  \, (-i\slashed{p})\, {s'}_{\pm}\right]}\ , \nonumber \\
             \left[   a_{\pm}(\textbf{p},s_{\pm},\sigma)  ,   a^{*}_{\pm}(\textbf{p}',s_{\pm}',\sigma')\right\rbrace  & \, = \,  \pm 2m\delta^{3}\left( \textbf{p}-\textbf{p}'\right) \delta_{\sigma\sigma'}\delta^{2}\left[ \left(  s- s'\right)_{\pm} \right] \ .
    \label{Massive_NonZeroMutations_SuperCreationAnnihilationOperators}
\end{IEEEeqnarray}
The $(+) $ and $ (-) $  creation-annihilation  sparticle operators are not independent; they are related by a Fourier transformation in fermionic variables. For creation type, we have
   \begin{IEEEeqnarray}{rl}
    a^{*}_{\pm}(\textbf{p},s_{\pm},\sigma)   \, = \,  \mp \left( 2m\right)^{-1} \int \,d^{2}s'_{\mp}\, \exp{\left[  2 s^{\intercal}_{\pm}\epsilon \gamma_{5} (+i\slashed{p})s'_{\mp} \right] } a^{*}_{\mp}(\textbf{p},s'_{\mp},\sigma) \ ,
     \label{Massive_CreationOperators_PlusMinus_AsMinusPlusIntegrals}
\end{IEEEeqnarray} 
and meanwhile for the annihilation type,
   \begin{IEEEeqnarray}{rl}
    a_{\pm}(\textbf{p},s_{\pm},\sigma)   \, = \,  \mp\left( 2m\right)^{-1} \int \,d^{2}s'_{\mp}\, \exp{\left[ - 2 s^{\intercal}_{\pm}\epsilon \gamma_{5} (+i\slashed{p})s'_{\mp} \right] } a_{\mp}(\textbf{p},s'_{\mp},\sigma) \ .
     \label{Massive_AnnihilationOperators_PlusMinus_AsMinusPlusIntegrals}
\end{IEEEeqnarray} 

Now, we introduce the Majorana  fermionic operators,
\begin{IEEEeqnarray}{l}
            \mathsf{U}(\Lambda)\mathcal{Q}_{\alpha}\mathsf{U}^{-1}(\Lambda)  \, = \,\sum_{\beta} D(\Lambda^{-1})_{\alpha\beta}{\mathcal{Q}}_{\beta}\ ,   \nonumber \\
            \left\lbrace \mathcal{Q}_{\alpha},\overline{\mathcal{Q}}_{\beta}\right\rbrace  \, = \, (-2i)\left( \gamma^{\mu}\right) _{\alpha\beta}\,\mathrm{P}_{\mu}\ ,  \quad [\mathcal{Q}_{\alpha},\mathrm{P}^{\mu}] \, = \, 0\ , 
    \label{SUSY_generators}
\end{IEEEeqnarray}
that are supersymmetry generators. We define a supersymmetric transformation through the exponential mapping
\begin{IEEEeqnarray}{l}
            \mathsf{U}(\vartheta) \, = \,  \exp{\left[+i{\vartheta}^{\intercal}\epsilon\gamma_{5}\mathcal{Q} \right] }\ ,
    \label{UnitarySUSYTransInFourComp}
\end{IEEEeqnarray}
 where $ \vartheta $ is a fermionic 4-spinor that parametrizes the transformation. The composition rule for the supersymmetric transformation is given by
\begin{IEEEeqnarray}{rl}
            \mathsf{U}(\vartheta')\mathsf{U}(\vartheta) & \, = \, \exp{\left[{i}\vartheta'^{\intercal} \epsilon\gamma_{5}\slashed{\mathrm{P}} \vartheta\right] }  \mathsf{U}(\vartheta+\vartheta') \ .
    \label{CompositionPropertySUSY}
\end{IEEEeqnarray}
 We take the  action  of  a supersymmetric transformation on creation-annihilation sparticle operators as
\begin{IEEEeqnarray}{rl}
             \mathsf{U}(\vartheta) a^{*}_{\pm}(\textbf{p},s_{\pm},\sigma)  \mathsf{U}(\vartheta)^{-1}  &  \, = \,    \exp{\left[\vartheta^{\intercal}\epsilon\gamma_{5}(+ i\slashed{p})\left(  2 s \, + \, \vartheta\right) _{\pm} \right]  } a^{*}_{\pm}(\textbf{p}, (s \, + \, \vartheta)_{\pm},\sigma)\ ,  \nonumber \\
             \mathsf{U}(\vartheta) a_{\pm}(\textbf{p},s_{\pm},\sigma)  \mathsf{U}(\vartheta)^{-1}  &  \, = \,    \exp{\left[\left( 2 {s} \, + \,  \vartheta\right)^{\intercal} \epsilon\gamma_{5}(+i\slashed{p})\vartheta_{\mp} \right]  }  a_{\pm}(\textbf{p}, (s \, +\, \vartheta)_{\pm},\sigma)\ . \nonumber \\
    \label{Massive_SUSYFinite_CreationAnnihilationOperators}
\end{IEEEeqnarray}

This equation is consistent with the composition property~\eqref{CompositionPropertySUSY}, with \eqref{Massive_CreationOperators_PlusMinus_AsMinusPlusIntegrals}, and \eqref{Massive_AnnihilationOperators_PlusMinus_AsMinusPlusIntegrals}. From here, we can write the finite supersymmetric transformations in components:
\begin{IEEEeqnarray}{rl}
              \mathsf{U}(\vartheta) a^{*}_{+}(\textbf{p},\sigma)  \mathsf{U}(\vartheta)^{-1}    &  \, = \,      \left[ 1 \, - \,m^{2}\delta^{4}(\vartheta)    \right]  a^{*}_{+}(\textbf{p},\sigma)   \, + \, \sqrt{{2}{m}} \,\overline{  b}(\textbf{p},\sigma) \,   \left[ {\vartheta}_{+} + m\delta^{2}(\vartheta_{+})\,  \vartheta_{-}\right] \nonumber \\
              &  \quad \, + \,   \left[  \vartheta^{\intercal}\epsilon\gamma_{5}(+ i\slashed{p})  \vartheta_{+}\, + \,  \, 2 m  \,\delta^{2}(\vartheta_{+})  \right] a^{*}_{+}(\textbf{p},\sigma)\ ,    \nonumber \\
              \mathsf{U}(\vartheta) a^{*}_{-}(\textbf{p},\sigma)  \mathsf{U}(\vartheta)^{-1}    &  \, = \,      \left[ 1 \, - \,m^{2}\delta^{4}(\vartheta)    \right]  a^{*}_{-}(\textbf{p},\sigma)   \, - \, \sqrt{{2}{m}} \,\overline{  b}(\textbf{p},\sigma) \,   \left[ {\vartheta}_{-} - m\delta^{2}(\vartheta_{-})\,  \vartheta_{+}\right] \nonumber \\
              &  \quad \, + \,   \left[  \vartheta^{\intercal}\epsilon\gamma_{5}(+ i\slashed{p})  \vartheta_{-}\, - \,  \, 2 m  \,\delta^{2}(\vartheta_{-})  \right] a^{*}_{-}(\textbf{p},\sigma)\ ,    \nonumber \\
              \mathsf{U}(\vartheta) \,\overline{  b}(\textbf{p},\sigma)   \mathsf{U}(\vartheta)^{-1}   &   \, = \,               \, + \,  \,\overline{  b}(\textbf{p},\sigma)  \left\lbrace I    \, + \,    m^{2}\delta^{4}(\vartheta)    \, + \,  {\tfrac{1}{4}}\left[ \vartheta^{\intercal} \epsilon\gamma_{\mu}    \vartheta\right] \gamma^{\mu} \left[ 
          m + i\slashed{p}     \right]\gamma_{5}  \right\rbrace \nonumber \\
                       \, + \,   \sqrt{\tfrac{m}{  2 }}   \,\left\lbrace  \left( {\tfrac{1}{  m }} \right. \right. &\left. \left.  \, + \, \,   \delta^{2}(\vartheta_{+}) \right) \, a^{*}_{- }(\textbf{p},\sigma) \, + \, \left( {\tfrac{1}{  m }}  \, - \, \,   \delta^{2}(\vartheta_{-}) \right) \, a^{*}_{+ }(\textbf{p},\sigma) \right\rbrace  \vartheta^{\intercal}\epsilon\left[ 
              m \, -\, i\slashed{p} \right]   \nonumber \\
                              \, + \,    \sqrt{\tfrac{m}{  2 }}   \,\left\lbrace  \left( {\tfrac{1}{  m }}\right.  \right. &\left. \left.  \, -\, \,   \delta^{2}(\vartheta_{+}) \right) \, a^{*}_{- }(\textbf{p},\sigma) \, -\, \left( {\tfrac{1}{  m }}  \, + \, \,   \delta^{2}(\vartheta_{-}) \right) \, a^{*}_{+ }(\textbf{p},\sigma) \right\rbrace  \vartheta^{\intercal}\epsilon\left[ 
              m \, +\, i\slashed{p} \right]\gamma_{5}\ .   \nonumber \\                                           
    \label{Massive_FiniteSUSYtransInComponents}
\end{IEEEeqnarray} 
We note that $ \mathsf{U}(\vartheta) \,\overline{  b(\textbf{p},\sigma)}   \mathsf{U}(\vartheta)^{-1}   $ is consistent with \eqref{MassivePropertyOf_b_DiracType}. Taking $ \vartheta $ infinitesimal,  Eq.  \eqref{Massive_FiniteSUSYtransInComponents} give us the following (anti)commutation relations:
\begin{IEEEeqnarray}{rl}       
 i      \left[a^{*}_{+}(\textbf{p},\sigma), {\mathcal{Q}}_{\alpha}\right\rbrace      &  \, = \,  +  \left(2m \right) ^{+1/2}  \,\left[ \overline{  b_{-}(\textbf{p},\sigma)}\epsilon\gamma_{5}\right]_{\alpha} \ ,   \nonumber \\  
  i      \left[a^{*}_{-}(\textbf{p},\sigma), {\mathcal{Q}}_{\alpha}\right\rbrace      &  \, = \,  -  \left(2m \right) ^{+1/2} \,\left[ \overline{  b_{+}(\textbf{p},\sigma)}\epsilon\gamma_{5}\right]_{\alpha} \ ,   \nonumber \\                                          
                  i\left\lbrace \overline{  b_{\alpha}}(\textbf{p},\sigma) ,{\mathcal{Q}}_{+\delta}\right]      &\, = \,+  \left(2m \right) ^{-1/2} \,a^{*}_{- }(\textbf{p},\sigma)  \left[ \left(I+\gamma_{5} \right) \left( 
              m \, -\, i\slashed{p} \right)\right]_{\delta\alpha}  \ ,   \nonumber \\
                           i\left\lbrace \overline{  b_{\alpha}}(\textbf{p},\sigma) ,{\mathcal{Q}}_{-\delta}\right]    & \, = \,     \, - \, \left(2m \right) ^{-1/2}\,  a^{*}_{+}(\textbf{p},\sigma)  \left[ \left(I-\gamma_{5} \right) \left( 
              m \, -\, i\slashed{p} \right)\right]_{\delta\alpha} \ .  \nonumber \\
    \label{SUSYInfinitesimal}
\end{IEEEeqnarray}
In the rest frame  $ L(k)=I $, therefore
\begin{IEEEeqnarray}{rl}
        i      \left[a^{*}_{+}(\textbf{k},\sigma), {\mathcal{Q}}_{a}\right\rbrace       & \, = \,     0\ ,  \qquad  i      \left[a^{*}_{-}(\textbf{k},\sigma), {\mathcal{Q}}^{*}_{a}\right\rbrace        \, = \,     0 \ , \nonumber \\
                    i      \left[a^{*}_{+}(\textbf{k},\sigma), {\mathcal{Q}}^{*}_{a}\right\rbrace   &     \, = \,     -\sqrt{{2}{m}} \,l^{*}_{a}(\textbf{k},\sigma)\ ,  \quad  i      \left[a^{*}_{-}(\textbf{k},\sigma), {\mathcal{Q}}_{a}\right\rbrace       \, = \,     \sqrt{{2}{m}} \,l^{*}_{b}(\textbf{k},\sigma)e_{ba}\ , \nonumber  \\
                 i\left\lbrace   l^{*}_{a}(\textbf{k},\sigma) ,{\mathcal{Q}}^{*}_{b}\right]   &  \, = \,   \sqrt{2m} a^{*}_{- }(\textbf{k},\sigma)  e_{ab}\ , \quad  i\left\lbrace   l^{*}_{a}(\textbf{k},\sigma) ,{\mathcal{Q}}_{b}\right]     \, = \,- \sqrt{2m}  a^{*}_{+}(\textbf{k},\sigma)  \delta_{ab}\ , \nonumber \\
    \label{SUSYmutatorAtRest}
\end{IEEEeqnarray}
recovering the structure of laddering operators of the fermionic generators (with steps $ \pm 1/2 $ in angular momentum).   Equations  \eqref{Massive_PoincareTrans_SuperCreationOperator} and \eqref{Massive_SUSYFinite_CreationAnnihilationOperators} show that under the  super Poincar\'e group  $   \mathsf{U}(\Lambda,x,\vartheta)   \, \equiv\,  \mathsf{U}(\Lambda,x)\mathsf{U}(\vartheta) $,
\begin{IEEEeqnarray}{rl}
         \mathsf{U}(\Lambda,x,\vartheta)   \left[         a_{\mp}(\textbf{p},s_{\mp},\sigma)   ,   a^{*}_{\pm}(\textbf{p}',s_{\pm}',\sigma')\right\rbrace \mathsf{U}(\Lambda,x,\vartheta)  ^{-1}   \, = \,\left[      a_{\mp}(\textbf{p},s_{\mp},\sigma)   ,   a^{*}_{\pm}(\textbf{p}',s_{\pm}',\sigma')\right\rbrace;  \nonumber \\
    \label{Massive_SuperPoincareTrans_BraketPreserving}
\end{IEEEeqnarray}
that is, the (anti)commutator of creation-annihilation sparticle operators remains invariant under a super Poncair\'e transformation. When  $ \vartheta $ satisfies the Majorana condition  $ \vartheta = \epsilon\gamma_{5}\beta\vartheta^{*} $,  Eq.  \eqref{Massive_SuperPoincareTrans_BraketPreserving} allows us to write  $   \left(  \mathsf{U}(\Lambda,x,\vartheta)^{-1} \right)^{*} =  \mathsf{U}(\Lambda,x,\vartheta)$ consistently. In other words, the sparticle state
\begin{IEEEeqnarray}{rl}
             \ket{\textbf{p},s_{\pm},\sigma}^{\pm}\, \equiv \,a^{*}_{\pm}(\textbf{p},s_{\pm},\sigma)\ket{\text{VAC}}
    \label{Massive_Superstates}
\end{IEEEeqnarray}
transforms unitarily  under the  super Poincar\'e group. Note also that
\begin{IEEEeqnarray}{rl}
               \mathsf{U}(\Lambda,x,\vartheta)    \left[        a_{\mp}(\textbf{p},s_{\mp},\sigma)  ,   a^{*}_{\mp}(\textbf{p}',s_{\mp}',\sigma')\right\rbrace       \mathsf{U}(\Lambda,x,\vartheta) ^{-1}  \, = \,  \left[    a_{\mp}(\textbf{p},s_{\mp},\sigma)  ,   a^{*}_{\mp}(\textbf{p}',s_{\mp}',\sigma')\right\rbrace \ . \nonumber \\
    \label{Massive_SUSYFinite_MutatorSoperator_plusminus_minusplus}
\end{IEEEeqnarray}

It is also possible to eliminate the quadratic phase factor appearing in~\eqref{Massive_SUSYFinite_CreationAnnihilationOperators}  by defining
\begin{IEEEeqnarray}{rl}
            a^{*}_{\pm}(\textbf{p},s,\sigma)   &   \,\equiv \, \exp{\left[\,s^{\intercal}\epsilon\gamma_{5}(-i\slashed{p})s_{\mp}\right] } a^{*}_{\pm}(\textbf{p},s_{\pm},\sigma) \ , \nonumber \\
              a_{\mp}(\textbf{p},s,\sigma)   &   \,\equiv \, \left(      a^{*}_{\pm}(\textbf{p},\epsilon\gamma_{5}\beta s^{*},\sigma)  \right)^{*} \  ,
    \label{Soperators_plusminus_Extended}
\end{IEEEeqnarray}
leading to
\begin{IEEEeqnarray}{rl}
               \mathsf{U}(\Lambda,x) a^{*}_{\pm}(\mathbf{p},s,\sigma)  \mathsf{U}(\Lambda,x)^{-1}&\, = \, e^{-i{\Lambda} p\cdot x} {\sqrt{\tfrac{\left(\Lambda p\right) ^{0}}{p^{0}}}}\sum_{\sigma'}U^{(j)}_{\sigma'\sigma}[W(\Lambda,\mathbf{p})]  a^{*}_{\pm}({\mathbf{p}_{\Lambda}},D(\Lambda)s,\sigma')\ ,  \nonumber \\
                   \mathsf{U}(\Lambda,x) a_{\pm}(\mathbf{p},s,\sigma)  \mathsf{U}(\Lambda,x)^{-1}&\, = \, e^{+ i {\Lambda} p\cdot x} {\sqrt{\tfrac{\left(\Lambda p\right) ^{0}}{p^{0}}}}\sum_{\sigma'}U^{(j) *}_{\sigma'\sigma}[W(\Lambda,\mathbf{p})]  a_{\pm}({\mathbf{p}_{\Lambda}},D(\Lambda)s,\sigma')\ ,  \nonumber \\
                   \mathsf{U}(\vartheta) a^{*}_{\pm}(\mathbf{p},s,\sigma)  \mathsf{U}(\vartheta)^{-1}  &  \, = \,    \exp{\left[\vartheta^{\intercal}\epsilon\gamma_{5}(+ i\slashed{p}) s  \right]  } a^{*}_{\pm}(\mathbf{p}, s \, + \, \vartheta,\sigma)   \ ,\nonumber \\
         \mathsf{U}(\vartheta)  a_{\pm}(\mathbf{p},s,\sigma)  \mathsf{U}(\vartheta)^{-1}  &  \, = \,    \exp{\left[ \vartheta^{\intercal}\epsilon\gamma_{5}(- i\slashed{p}) s  \right]  }  a_{\pm}(\mathbf{p}, s \, +\, \vartheta,\sigma)   \ .
     \label{SuperState_PoincareTrans_Extended}
\end{IEEEeqnarray}

\section{Causal Superfields.}\label{Sec_CausaleSuperfields}
Now, we are in a position to define causal quantum superfields out of momentum superspace Fourier transformations of the creation-annihilation sparticle operators. We choose supersymmetric transformations in configuration superspace that induce linear-homogeneous ones in the spacetime variable $ x^{\mu} $, and they in turn generate symmetric covariant superderivatives~\cite{Srivastava:1974zn}. It has to be noted that in
this formalism these superderivatives arise directly from considering the most general superfield, without any other extra input. As in ordinary quantum field theory, we introduce two kinds of superfields,
	\begin{IEEEeqnarray}{l}             
                \Xi^{*}_{\pm n}( x,\vartheta)  \, \equiv \,    \sum_{\sigma}\int d^{3}\textbf{p}\, \, d^{4}s\,  {a}^{*}_{\pm}(\textbf{p},s,\sigma) v_{\pm {n}}(x,\vartheta;\textbf{p},s,\sigma)  \ ,  
    \label{Creation_Annihilation_Sfield_Definition-1}
 \\
                 \Xi_{\pm {n}}( x,\vartheta)  \, \equiv \, \sum_{\sigma} \int d^{3}\textbf{p}\,  \, d^{4}s \, {a}_{\pm}(\textbf{p},s,\sigma)  u_{\pm {n}}(x,\vartheta;\textbf{p},s,\sigma) \ ,  
    \label{Creation_Annihilation_Sfield_Definition-2}
\end{IEEEeqnarray}
that give a total of four superfields. The quantities $ u_{\pm {n}} $  and  $ v_{\pm {n}} $ are the corresponding super wave functions that are determined by demanding  for    $ \Xi^{*}_{\pm {n}} $ the  super Poincar\'e transformation,
               \begin{IEEEeqnarray}{rl}
                              \mathsf{U}(\Lambda,a) \Xi^{*}_{\pm {n}}(x,\vartheta) \mathsf{U}(\Lambda,a) ^{-1}  &\, = \, \sum_{  \pm{m}}\left[ {S}(\Lambda^{-1})\right] _{ \pm{n},\pm {m}}{\Xi}^{*}_{\pm {m}}\left( \Lambda x+a ,D(\Lambda)\vartheta\right) \ , \\                         
                                   \mathsf{U}(\xi) {\Xi}^{*}_{\pm {n}}(x,\vartheta) \mathsf{U}(\xi) ^{-1}  &\, = \, {\Xi}^{*}_{\pm {n}}\left(  x^{\mu}+\vartheta^{\intercal}\epsilon\gamma_{5}\gamma^{\mu}\xi,\vartheta \, + \, \xi\right)  \ ,                         
       \label{TransformationRuleOfTheFields_SuperPoincare}
                  \end{IEEEeqnarray} 
where ${S}_{\pm{n},\pm{m}} $ is a finite-dimensional Lorentz representation that in principle could be different for $ \Xi^{*}_{+{n}}  $ and $ \Xi^{*}_{- {n}} $. With the help of~\eqref{SuperState_PoincareTrans_Extended}, the general solution of~\eqref{Creation_Annihilation_Sfield_Definition-1}, and including the requirements in~\eqref{TransformationRuleOfTheFields_SuperPoincare}, can be expressed as
 	\begin{IEEEeqnarray}{ll}
          {\Xi}^{*}_{\pm {n}}(x,\vartheta)  &\, = \,   \sum_{\sigma}  \int d^{3}\textbf{p}\,d^{4}{s}\,\,e^{-i x\cdot p}   e^{ {\vartheta}^{\intercal}\epsilon\gamma_{5}(+i\slashed{p}){s}}\,{a}^{*}_{\pm}(\mathbf{p},{s},\sigma)     {v}_{\pm {n}}\left( \textbf{p}, (-i\slashed{p})\left[ {s}  -{\vartheta}\right] ,\sigma\right)  \ . \nonumber \\
     \label{Non_Causal_General_SuperPoncaireCovSField}
 \end{IEEEeqnarray}
 The coefficients  $  {v}_{\pm {n}}\left( \textbf{p}, (-i\slashed{p})\left[ {s}  -{\vartheta}\right] ,\sigma\right)  $ are given in the rest frame: 
  	 	\begin{IEEEeqnarray}{rl}                                
             {v}_{\pm {n}} &\left( \textbf{p}, (-i\slashed{p})\left[ {s}  -{\vartheta}\right] ,\sigma\right)  \nonumber \\
                     \, = \,   \sqrt{\tfrac{k^{0}}{p^{0}}} \sum_{ \pm{m}}   &  \left[ {S}(L(p))\right] _{ \pm{n},\pm {m}}{v}_{\pm {n}}\left( \textbf{k}, (-i\slashed{k})D[L(p)]^{-1}\left[ {s}  -{\vartheta}\right],\sigma\right) \ .        
             \label{MatchingLorentzFourirerCoefficientsFromStandardVector}
    	\end{IEEEeqnarray} 
Given a unitary representation for the superstate of superspin $ j $, the coefficients in the rest frame are required to satisfy
 	\begin{IEEEeqnarray}{rl}                
         \sum_{\sigma'}     {v}_{\pm {n}} &\left( \textbf{k}, (-i\slashed{k})\left[ {s}  -{\vartheta}\right],\sigma'\right) U^{(j)*}_{\sigma'\sigma }(W) \nonumber \\
          \, = \,  & \sum_{ \pm{m}}  \left[ {S}(W)\right] _{ \pm{n},\pm {m}} {v}_{\pm {n}} \left( \textbf{k}, (-i\slashed{k})D\left[ W^{-1}\right] \left[ {s}  -{\vartheta}\right],\sigma\right) \ ,
             \label{MatchingLorentzFourirerCoefficientsLittleGroup}
    	\end{IEEEeqnarray}
     	with $ W $ being a little group transformation of the form~\eqref{WignerTransformation}.  Equations \eqref{MatchingLorentzFourirerCoefficientsFromStandardVector} and \eqref{MatchingLorentzFourirerCoefficientsLittleGroup} have to be satisfied by the expansion coefficients of the $ \vartheta-s $ variables independently, showing that the superfield \eqref{Non_Causal_General_SuperPoncaireCovSField} is a reducible realization of the  super Poincar\'e symmetry.  

Consider the zero  order fermionic expansion in $ v_{\pm n} $ for the annihilation superfield:
\begin{IEEEeqnarray}{ll}
          {\chi}^{*}_{\pm {n}}(x,\vartheta)  &\, \equiv \,   -\frac{1}{m^{2}}\sum_{\sigma}  \int d^{3}\textbf{p}\,d^{4}{s}\,\,e^{-i x\cdot p}   e^{ {\vartheta}^{\intercal}\epsilon\gamma_{5}(+i\slashed{p}){s}}\,{a}^{*}_{\pm}(\mathbf{p},{s},\sigma)    v_{\pm {n}}(\textbf{p} ,\sigma)   \ .    
     \label{Chiral_Creation_SuperFields1}
 \end{IEEEeqnarray}
Since we can generate  terms of the form $ \left[ \slashed{p}(\vartheta-s)\right]_{\alpha}  $ by applying the  superderivative defined as
     	\begin{IEEEeqnarray}{l}
     	 \mathcal{D}   \, \equiv \,    \left(\epsilon\gamma_{5} \right)\frac{\partial}{\partial{\vartheta} } \, - \,  \gamma^{\mu}\vartheta\frac{\partial}{\partial x^{\mu}} \ ,
    \label{CovariantDerivativeExplicitely}
\end{IEEEeqnarray} 
we can reconstruct the reducible superfields $ {\Xi}^{*}_{\pm {n}}(x,\vartheta)  $  from superfields of the form~\eqref{Chiral_Creation_SuperFields1}. 
We can also  introduce a  zero order creation superfield   $   {\chi}_{\pm {n}}(x,\vartheta)  $:
\begin{IEEEeqnarray}{ll}
          {\chi}_{\pm {n}}(x,\vartheta)  &\, \equiv \,   -\frac{1}{m^{2}}\sum_{\sigma}  \int d^{3}\textbf{p}\,d^{4}{s}\,\,e^{+i x\cdot p}   e^{ {\vartheta}^{\intercal}\epsilon\gamma_{5}(-i\slashed{p}){s}}\,{a}_{\pm}(\mathbf{p},{s},\sigma)    u_{\pm {n}}(\textbf{p} ,\sigma)    \ .   
     \label{Chiral_Annihilation_SuperFields1}
 \end{IEEEeqnarray}

Given $n=(a,b)$, where  $a =-\mathcal{A} , -\mathcal{A} +1, \dots, \mathcal{A} -1, \mathcal{A}$ and $b =-\mathcal{B} , -\mathcal{B} +1, \dots, \mathcal{B} -1,\mathcal{B}$, and $ 2\mathcal{A},2\mathcal{B}  \, = \,0, 1,2,\dots$,
 we enumerate irreducible finite representations of the Lorentz group  by the $SU(2) $ pair of indices $ (\mathcal{A},\mathcal{B}) $.

 Depending on whether we operate an even or odd number of times the $  \mathcal{D} $'s, we obtain all the possible superspins that an irreducible representation $ S_{\pm m \pm n} $  can carry. For the zero order and the first superderivative  we have
\begin{IEEEeqnarray}{rlr}
             \vert \mathcal{A}-\mathcal{B}\vert \leq & \,\, j \, \leq\vert \mathcal{A}+\mathcal{B} \vert ,  &\text{zero order in }\, \mathcal{D}_{\alpha}  \ ;
             \label{InequalitiesForSuperspinReal}     \\             
      \vert \mathcal{A}-\mathcal{B}\pm \tfrac{1}{2}\vert  \leq  &  \,\, j  \,\leq\vert \mathcal{A}+\mathcal{B}\pm  \tfrac{1}{2}\vert ,& \quad \text{linear in }\, \mathcal{D}_{\alpha}  \ .
    \label{InequalitiesForFakeSuperspinFake}
\end{IEEEeqnarray}
These relations follow  from \eqref{MatchingLorentzFourirerCoefficientsLittleGroup} and the product rules of $ \left( \mathcal{A},\mathcal{B}\right)\otimes\left[ \left(\frac{1}{2},0 \right)\oplus \left(0,\frac{1}{2} \right) \right]  $.  With the help of Eq.~\eqref{Soperators_plusminus_Extended}, we can integrate  explicitly the superfields \eqref{Chiral_Creation_SuperFields1}  and \eqref{Chiral_Annihilation_SuperFields1} in the fermionic variable $ s $  to obtain
\begin{IEEEeqnarray}{ll}
          {\chi}^{*}_{\pm {n}}(x,\vartheta)  
           &\, = \,  \sum_{\sigma}  \int d^{3}\textbf{p}\,e^{ -i x_{\pm}\cdot p  } \,{a}^{*}_{\pm}(\mathbf{p},{\vartheta}_{\pm},\sigma)  v_{{n}}(\textbf{p} ,\sigma)        \ ,  
     \label{Chiral_Creation_SuperFields}
 \\
         {\chi}_{\pm {n}}(x,\vartheta)         
           &\, = \,  \sum_{\sigma}  \int d^{3}\textbf{p}\,e^{ +i x_{\pm}\cdot p  } \,{a}_{\pm}(\mathbf{p},{\vartheta}_{\pm},\sigma)  u_{{n}}(\textbf{p} ,\sigma)      \ ,   
     \label{Chiral_Annhihilation_SuperFields}
 \end{IEEEeqnarray} 
 where $ x^{\mu}_{\pm}  \, = \, x^{\mu} \, - \,\vartheta^{\intercal}\epsilon\gamma_{5}\gamma^{\mu}\vartheta _{\pm} $. Note that in Eqs. \eqref{Chiral_Creation_SuperFields} and \eqref{Chiral_Annhihilation_SuperFields} we are dropping the sign $ {\pm} $ in the Fourier coefficients $ u_{{n}} $  and $ v_{{n}} $ because the inequalities \eqref{InequalitiesForSuperspinReal} and \eqref{InequalitiesForFakeSuperspinFake} allow us to consider $\pm $ superfields for one and the same representation. From now on,  we will suppose that this is case. We can see that these zero order superfields  are chiral,
\begin{IEEEeqnarray}{rl}
              \mathcal{D}_{\mp}  \begin{pmatrix}
              {\chi}^{*}_{\pm {n}}(x,\vartheta)  \\ 
               {\chi}_{\pm {n}}(x,\vartheta)  
              \end{pmatrix}   \, = \,  0 \ , 
    \label{Quiral_Conditions_NonCausalSfields}
\end{IEEEeqnarray}
and also that
\begin{IEEEeqnarray}{rl}
                 \mathcal{D}^{\intercal}_{\pm} \epsilon \mathcal{D}_{\pm} \begin{pmatrix}
              {\chi}^{*}_{\pm {n}}(x,\vartheta)  \\ 
               {\chi}_{\pm {n}}(x,\vartheta)  
              \end{pmatrix}  \, = \,  \mp 4 m \begin{pmatrix}
              {\chi}^{*}_{\mp {n}}(x,\vartheta)  \\ 
               {\chi}_{\mp {n}}(x,\vartheta)  
              \end{pmatrix}  \ .
    \label{MotionEquations_NonCausal}
\end{IEEEeqnarray}
The last set of equations are usually taken as the free equations of motion. For us, they mean we can work with  $ {\chi}_{+ {n}}(x,\vartheta) $ and  $ {\chi}_{-{n}}(x,\vartheta) $ without the need to introduce $ \mathcal{D}^{\intercal}_{\pm} \epsilon \mathcal{D}_{\pm}  $, or just work with (+) superfields $ {\chi}_{+ {n}}(x,\vartheta) $  and $ \mathcal{D}^{\intercal}_{+} \epsilon \mathcal{D}_{+}  {\chi}_{+ {n}}(x,\vartheta)$   (similar remarks for  $ \chi^{*}_{\pm n} $).  From the relation
 \begin{IEEEeqnarray}{rl}
           \left\lbrace  \mathcal{D}_{\alpha}, \mathcal{D}_{\beta} \right\rbrace              & \, = \,  \, + \, 2 \left(  \gamma^{\mu} \epsilon\gamma_{5}\right) _{\alpha\beta} \partial_{\mu}\   ,    \quad 
              \label{AnticommutoarOFSuperderivatives}     
\end{IEEEeqnarray}
and Eq. \eqref{Quiral_Conditions_NonCausalSfields},    $ p_{+}$ products of $ \mathcal{D}_{+\alpha} $ superderivatives together with $ p_{-}$  products of $ \mathcal{D}_{-\beta} $ superderivatives acting on  $ \chi_{\pm {n}}(x,\vartheta) $ are equivalent to   $ p_{\pm }$  products of $ \mathcal{D}_{\pm\alpha} $ acting on  $ \chi_{\pm {n}}(x,\vartheta)   $ plus sums  of $ p'_{\pm} < p_{\pm} $ products of $ \mathcal{D}_{\pm\alpha} $ times ordinary derivatives $ \partial_{\mu} $  acting on $ \chi_{\pm {n}}(x,\vartheta)   $. Also from \eqref{AnticommutoarOFSuperderivatives}, $ \left\lbrace\mathcal{D}_{\pm\alpha},\mathcal{D}_{\pm\beta} \right\rbrace =0  $, which means that nonzero  products of superderivatives of the same sign end at the second order  $  \mathcal{D}_{\pm\alpha}\mathcal{D}_{\pm\beta}$, but  $ \mathcal{D}_{\pm\alpha}\mathcal{D}_{\pm\beta} =  {\frac{1}{4}\left[ \epsilon\left( 1\, \pm \,\gamma_{5}\right)\right] _{\alpha\beta}} \left(\mathcal{D}_{\pm}^{\intercal}\epsilon \mathcal{D}_{\pm}\right)  $,  which due to~ \eqref{MotionEquations_NonCausal} flips the signs of $ \chi_{\pm {n}}(x,\vartheta)   $ to $ \chi_{\mp {n}}(x,\vartheta)   $ [same remarks for $ \chi^{*}_{\pm {n}}(x,\vartheta) $]. Finally, since derivatives of superfields can be taken as superfields without derivatives, with complete generality, we  can consider  superfields of the form~\footnote{ Expressions $\left(  \mathcal{D}{\chi}_{{n}} \right) _{\pm\alpha} $ and  $ \left( \mathcal{D}{\chi}^{*}_{{n}} \right) _{\pm\alpha} $ are shorthand  notations for $ \mathcal{D}_{\pm\alpha} {\chi}_{\pm{n}} $ and $  \mathcal{D}_{\pm\alpha} {\chi}^{*}_{\pm{n}} $, respectively.}
\begin{IEEEeqnarray}{rl}
      {\chi}_{\pm{n}}, \quad      {\chi}^{*}_{\pm{n}}, \quad \left( \mathcal{D}{\chi}_{{n}} \right) _{\pm\alpha} ,\quad  \left( \mathcal{D}{\chi}^{*}_{{n}} \right) _{\pm\alpha}   \ . 
    \label{ListOfPosibleSuperfields}
\end{IEEEeqnarray}
For a fixed irreducible representation of the Lorentz group, due to~\eqref{InequalitiesForSuperspinReal} and~\eqref{InequalitiesForFakeSuperspinFake}, chiral superfields and linear superderivatives  of chiral superfields are incompatible.
Now, we introduce causal superfields 
\begin{IEEEeqnarray}{rl}           
                \Phi_{\pm n}(x,\vartheta)        \, = \,       (2\pi)^{-3/2}\sum_{\sigma}  \int d^{3}\textbf{p}\, &\left\lbrace \,e^{ +i\left(  x_{\pm}\cdot p \right) }  {a}_{\pm}(\mathbf{p},{\vartheta}_{\pm},\sigma)  {u}_{n }(\textbf{p} ,\sigma) \right.  \nonumber \\
  &  \left.          \qquad   \, + \,(-)^{2\mathcal{B}}\, e^{ -i\left(  x_{\pm}\cdot p \right) } \,{a}^{c\,*}_{\pm}(\mathbf{p},{\vartheta}_{\pm},\sigma)    {v}_{n }(\textbf{p} ,\sigma)  \right\rbrace \ , \nonumber \\
   \Phi^{*}_{\pm n}(x,\vartheta)        \, = \,       (2\pi)^{-3/2}\sum_{\sigma}  \int d^{3}\textbf{p}\, &\left\lbrace \,(-)^{2\mathcal{B}} e^{ +i\left(  x_{\pm}\cdot p \right) }  {a}^{c}_{\pm}(\mathbf{p},{\vartheta}_{\pm},\sigma) \left( {v}_{n }(\textbf{p} ,\sigma) \right)^{*}  \right.  \nonumber \\
  &  \left.          \qquad   \, + \,\, e^{ -i\left(  x_{\pm}\cdot p \right) } \,{a}^{*}_{\pm}(\mathbf{p},{\vartheta}_{\pm},\sigma)    \left( {u}_{n }(\textbf{p} ,\sigma)\right)^{*}   \right\rbrace \ , \nonumber \\
    \label{CausalLeft_Right_Chiral-Sfields}
\end{IEEEeqnarray}   
with $ {v}_{n} (\textbf{p},\sigma)  \, = \,(-)^{j+\sigma} {u}_{ n} (\textbf{p},-\sigma) $ (for explicit formulas of these wave functions see Ref.~\cite{Weinberg:1969di}). Note that they are related by 
\begin{IEEEeqnarray}{rl}
               \Phi^{*}_{\mp n}(x,\vartheta) = \left( \Phi_{\pm n}(x,\epsilon\gamma_{5}\beta\vartheta^{*}) \right) ^{*} \ .
     \label{CauselSuperfieldStarred}
 \end{IEEEeqnarray} 
 Consider now another superfield $ \tilde{\Phi}^{*}_{\mp\tilde{n}}(x',\vartheta')  $ for the same sparticle. Introducing 
\begin{IEEEeqnarray}{rl}
            \left(  x^{\pm }_{_{12}}\right)^{\mu}   \, = \, x^{\mu}_{_{1}}-x^{\mu}_{_{2}}  \, + \, (\vartheta_{_{2}}-\vartheta_{_{1}})^{\intercal}\epsilon\gamma_{5}\gamma^{\mu}(\vartheta_{_{2}\mp} +\vartheta_{_{1}\pm})   \, = \, -\left(  x^{\mp }_{_{21}}\right)^{\mu} \ ,
    \label{x_plus_12}
\end{IEEEeqnarray}
 we can we write the (anti)commutator of $ \Phi_{\pm n}(x_{_{1}},\vartheta_{_{1}}) $ and  $ \tilde{\Phi}^{*}_{\mp\tilde{n}}(x_{_{1}},\vartheta_{_{2}})$ as
 \begin{IEEEeqnarray}{rl}
    \left[\Phi_{\pm n}(x_{_{1}},\vartheta_{_{1}}) , \tilde{\Phi}^{*}_{\mp\tilde{n}}(x_{_{2}},\vartheta_{_{2}})  \right]_{\varepsilon}         &\, = \,  (2\pi)^{-3}\int d^{3}\textbf{p}(2p^{0})^{-1}  \,\,\exp{\left[ +i x^{\pm}_{_{12}}\cdot p\right] }     {P}_{ n,\tilde{n}}  \left(  \textbf{p},p^{0}\right)  \nonumber \\
     \, + \, &   \varepsilon (-)^{2(\mathcal{B}+\tilde{\mathcal{B}})} (2\pi)^{-3} \int d^{3}\textbf{p}(2p^{0})^{-1}  \,\,\exp{\left[ -i x^{\pm}_{_{12}}\cdot p\right] }     {P}_{ n,\tilde{n}}  \left(  \textbf{p},p^{0}\right)\ ,  \nonumber \\
    \label{MutationRel_Causal_Fields_plusminus_plusminus}
\end{IEEEeqnarray}
with $ \varepsilon = -1 $ for commutator and $ \varepsilon = +1 $ for anticommutator.    $ {P}_{ n,\tilde{n}}  \left(  \textbf{p},p^{0}\right) $ can be expressed as\cite{Weinberg:1969di}
\begin{IEEEeqnarray}{rl}
              {P}_{ n,\tilde{n}}  \left(  \textbf{p},p^{0}\right)   \, = \,   {P}_{ n,\tilde{n}}  \left(  \textbf{p}\right)   \, + \,  p^{0}\,{Q}_{ n,\tilde{n}}  \left(  \textbf{p}\right) \ ,
    \label{LinearizedMomentumPropagator}
\end{IEEEeqnarray}
where $   {P}_{ n,\tilde{n}}  \left(  \textbf{p}\right)  $ and  $ {Q}_{ n,\tilde{n}}  \left(  \textbf{p}\right)  $   polynomials in  $ \mathbf{p} $ obtained from 
\begin{IEEEeqnarray}{rl}
            (2p^{0})^{-1} {P}_{ n,\tilde{n}}  \left(  \textbf{p},p^{0}\right)  &\, = \,  \sum_{\sigma} {u}_{n}(\textbf{p},\sigma)\tilde{u}^{*}_{\tilde{n}}(\textbf{p},\sigma)  \, = \, \sum_{\sigma} {v}_{n}(\textbf{p},\sigma)\tilde{v}^{*}_{\tilde{n}}(\textbf{p},\sigma) \ .\nonumber \\
    \label{Ordinary_Momentum_PrePropagator}
\end{IEEEeqnarray} 
Weinberg has shown~\cite{Weinberg:1969di} that $ P_{ n,\tilde{n}}    (\textbf{p},p^{0})   \, = \, (-)^{2(\mathcal{A}+\tilde{\mathcal{B}})}  P_{ n,\tilde{n}}    (-\textbf{p},-p^{0})  $,   
 therefore at $ (x_{_{1}}-x_{_{2}})^{2}>0 $,
\begin{IEEEeqnarray}{rl}
   \left[\Phi_{\pm n}(x_{_{1}},\vartheta_{_{1}}) , \tilde{\Phi}^{*}_{\mp\tilde{n}}(x_{_{2}},\vartheta_{_{2}})  \right]_{\varepsilon}       \, = \,  
    &\left( 1 \, + \,  \epsilon (-)^{2(\mathcal{A}\, + \, {\mathcal{B}})} \,   \right)  {P}_{n,\tilde{n}}    \left(-i\partial_{_{1}}\right) \Delta_{+}(x^{\pm}_{_{12}})\ , 
    \label{MutationRel_Causal_At_Spacelike}
\end{IEEEeqnarray}  
with
\begin{IEEEeqnarray}{rl}
             \Delta_{+}(x^{\pm}_{_{12}})   \, = \,       (2\pi)^{-3}    \int d^{3}\,\textbf{p} \,(2p^{0})^{-1}  \,\exp{\left[ +i x^{\pm}_{_{12}}\cdot p\right] }     \ .
    \label{DeltaPlus}
\end{IEEEeqnarray}
  Equation \eqref{MutationRel_Causal_At_Spacelike} vanishes  provided  that $ \varepsilon \, = \,  -(-)^{2(\mathcal{A}+\mathcal{B})}   \, = \, -(-)^{2j}   $. For linear superderivatives of chiral superfields, the vanishing of  the expression 
\begin{IEEEeqnarray}{rl}
             \left[\left(\mathcal{D} \Phi_{n'}(x_{_{1}},\vartheta_{_{1}}) \right)_{\pm\alpha} , (\mathcal{D} \tilde{\Phi}^{*}_{\tilde{n}'}(x_{_{2}},\vartheta_{_{2}}) )_{\mp\beta}  \right]_{\left( -\varepsilon'\right) }    
      \label{LinearDerivatives}
  \end{IEEEeqnarray}  
   at spacelike separations  gives $ \varepsilon' \, = \, -(-)^{2j}   \, = \, -\varepsilon  $, therefore making  $ \Phi_{\pm n}$  and  $ \left(\mathcal{D} \Phi_{n'}\right) _{\pm\alpha} $ incompatible. Since $ \Phi_{\pm n}$  goes in accordance with the spin statistics theorem, from now on \emph{we will leave out}  $ \left(\mathcal{D} \Phi_{n'}\right) _{\pm\alpha} $ from the discussion.  

Causal superfields are  also chiral,
\begin{IEEEeqnarray}{rl}
              \mathcal{D}_{\mp}  \begin{pmatrix}
              \Phi^{*}_{\pm {n}}(x,\vartheta)  \\ 
               \Phi_{\pm {n}}(x,\vartheta)  
              \end{pmatrix}   \, = \,  0\ , 
    \label{Quiral_Conditions_CausalSfields}
\end{IEEEeqnarray}
and satisfy
\begin{IEEEeqnarray}{rl}
                 \mathcal{D}^{\intercal}_{\pm} \epsilon \mathcal{D}_{\pm} \begin{pmatrix}
              \Phi^{*}_{\pm {n}}(x,\vartheta)  \\ 
               \Phi_{\pm {n}}(x,\vartheta)  
              \end{pmatrix}   \, = \,  \mp 4 m  \begin{pmatrix}
              \Phi^{*}_{\mp {n}}(x,\vartheta)  \\ 
               \Phi_{\mp {n}}(x,\vartheta)  
              \end{pmatrix}  \ .
    \label{MotionEquations_Causalsuperfields}
\end{IEEEeqnarray}

Expanding the superfields as
\begin{IEEEeqnarray}{rl}
                  \Phi_{\pm n}(x,\vartheta)    \, = \,  \phi_{\pm n}(x_{\pm})    \, \mp \,\sqrt{2} \vartheta_{\pm}^{\intercal}\epsilon\gamma_{5} \psi_{n}(x_{\pm})   \, \pm \,  2m \delta^{2}\left(\vartheta_{\pm}\right) \phi_{\mp n}(x_{\pm}) \ ,
    \label{SuperFieldInComponents}
\end{IEEEeqnarray}
we have
\begin{IEEEeqnarray}{rl}           
             \phi_{\pm n}(x)      \, = \,       (2\pi)^{-3/2}\sum_{\sigma}  \int d^{3}\textbf{p}\, &\left\lbrace \,e^{ +i  x\cdot p  }  {a}_{\mp}(\mathbf{p},\sigma)  {u}_{n }(\textbf{p} ,\sigma) \right.  \nonumber \\
  &  \left.          \qquad   \, + \,(-)^{2\mathcal{B}}\, e^{ -ix\cdot p } \,{a}^{c\,*}_{\pm}(\mathbf{p},\sigma)    {v}_{n }(\textbf{p} ,\sigma)  \right\rbrace  \  ,  \nonumber \\
    \left[ \psi_{n}(x) \right] _{\alpha}     \, = \,     \sqrt{m}  (2\pi)^{-3/2}\sum_{\sigma}  \int d^{3}\textbf{p}\, &\left\lbrace \,e^{ +i  x\cdot p  }\, \left[ b(\mathbf{p},\sigma)\right]_{\alpha}   {u}_{n }(\mathbf{p} ,\sigma) \right.  \nonumber \\
  &  \left.          \quad  \, - \,(-)^{2\mathcal{B}+2j}\, e^{ -ix\cdot p } \,  \left[ \epsilon \gamma_{5}\beta\, b^{c *}(\mathbf{p},\sigma)\right]_{\alpha}   {v}_{n }(\textbf{p} ,\sigma)  \right\rbrace   \ ,\nonumber \\
    \label{Components_CausalLeft_Right_Chiral-Sfields}
\end{IEEEeqnarray} 
with $ \psi_{n}(x) $ satisfying Dirac's equation: $ \left( \slashed{\partial} \, + \, m\right) \psi_{n}(x) =0 $. Now, it is clear that  one of the roles of the superfields $ \Phi_{-n} $ and $ \Phi_{+n} $   is to allow us to  use  $ \left(0,\frac{1}{2}\right)\otimes\left( \mathcal{A},\mathcal{B}\right)    $ and $  \left(\frac{1}{2},0\right)\otimes\left( \mathcal{A},\mathcal{B}\right)   $, respectively,  for their linear terms.  The component fields in \eqref{Components_CausalLeft_Right_Chiral-Sfields} satisfy  Klein–Gordon equations, since the $ \psi_{n} $  also satisfy the Dirac equation, the number of independent components $ \phi_{+n} $  and $ \phi_{-n} $ are equal to the number of independent components of   $ \psi_{n} $. There could be more redundancy equations that the three fields will share.

\section{Time-ordered products and superpropagators.}\label{Sec_TimeOrderedSuperpropagators}
So far, everything has gone as in  ordinary quantum field theory, but things are different for superpropagators: time-ordered products in Dyson-series are not supersymmetric invariant, and we need to correct  them in order to write superpropagators  properly. We start by writing  the superpropagator that follows from Wick's pairing rules,
\begin{IEEEeqnarray}{ll}
   -i\tilde{\Delta}^{\pm \mp}_{n,\tilde{n}}\left( x_{_{1}},\vartheta_{_{1}},x_{_{2}},\vartheta_{_{2}}\right)        
       &\, = \,  \omega(x_{_{12}}^{0})(2\pi)^{-3}  \,  {P}_{ n,\tilde{n}}  \left(  -i\frac{\partial}{\partial x_{_{1}}}\right)\Delta_{+}\left( x^{\pm}_{_{12}}\right)    \nonumber \\
   &  \, + \, \omega(x_{_{21}}^{0})(2\pi)^{-3} {P}_{ n,\tilde{n}}  \left(  -i\frac{\partial}{\partial x_{_{1}}}\right)\Delta_{+}\left( -x^{\pm}_{_{12}}\right)  \ ,
    \label{NonCorrectedSuperpropagator}
\end{IEEEeqnarray} 
where $ \omega(x_{_{12}}^{0})  \, = \, \omega(x_{_{1}}^{0}-x_{_{2}}^{0}) $ is the step function. To  illustrate that this superpropagator is not supersymmetric invariant, we consider interactions restricted to superpotentials\footnote{The use of $ \delta^{2}\left(\vartheta_{+} \right)\mathcal{W}_{+}\left(x,\vartheta \right) $ or   $ \delta^{2}\left(\vartheta_{-}\right) \mathcal{W}_{-}\left(x,\vartheta \right) $  in the first term of the superpotential is merely conventional since we can always make the redefinition $ \mathcal{W}'_{\pm}\left(x,\vartheta \right)=\mathcal{W}^{*}_{\pm}\left(x,\vartheta \right)$.  }
\begin{IEEEeqnarray}{rl}
            \mathcal{V}\left( x,\vartheta\right)    &\, = \,     \mathcal{V}_{\pm}\left(x,\vartheta\right)    \, + \,          \mathcal{V}^{*}_{\mp}\left(x,\vartheta\right)  \ ,\nonumber \\
     \mathcal{V}_{\pm}\left(x,\vartheta\right)     &  \, = \,      i\delta^{2}\left(\vartheta_{\mp} \right) \mathcal{W}_{\pm}\left(x,\vartheta \right)  \ ,
    \label{SuperPotentials}
\end{IEEEeqnarray}
where 
\begin{IEEEeqnarray}{rl}
              \mathcal{W}^{*}_{\pm}\left(x,\vartheta \right)   \, = \, \left( \mathcal{W}_{\mp}\left(x,\epsilon\gamma_{5}\beta\vartheta^{*} \right)\right)^{*} , \quad \mathcal{D}_{\mp}\mathcal{W}_{\pm}\left(x,\vartheta \right)=0  \ .
    \label{PropertiesOfSuperPotential}
\end{IEEEeqnarray}
  Its general component expansion can be expressed as
\begin{IEEEeqnarray}{l}
            \mathcal{W}_{\pm}(x,\vartheta)   \, = \, \mathcal{C}\left(x_{\pm} \right)   \, + \, \sqrt{2}\,\vartheta_{\pm}^{\intercal}\epsilon \,\Omega\left(x_{\pm} \right)   \, +  \, \delta^{2}(\vartheta_{\pm})\mathcal{F}\left(x_{\pm} \right)  \ .
    \label{SuperpotentialComponents}
\end{IEEEeqnarray}  
 Further restricting it to scalar superfields, the superpropagator then becomes (dropping the $ -i $ factor for now)
\begin{IEEEeqnarray}{rl}
  \delta^{2}(\vartheta_{_{1}\mp})\delta^{2}  (\vartheta_{_{2}\pm}) &\tilde{\Delta}^{\pm \mp}\left( x_{_{1}},\vartheta_{_{1}},x_{_{2}},\vartheta_{_{2}}\right)  \nonumber\\
       \, = \,  \delta^{2} & (\vartheta_{_{1}\mp})\delta^{2}  (\vartheta_{_{2}\pm}) \left[1  \, +\, 2 \vartheta_{_{1}}^{\intercal}\epsilon\gamma_{5}(-\slashed{\partial}_{_{1}})\vartheta_{_{2}\mp}  \, - \,4 m^{2}\,\delta^{2}(\vartheta_{_{1}\pm})\delta^{2}  (\vartheta_{_{2}\mp})  \right] \Delta_{F}(x_{_{1}}-x_{_{2}}) \ ,\nonumber \\  
    \label{MutationRel_Causal_Fields_plusminus_plusminus}
\end{IEEEeqnarray}
with
\begin{IEEEeqnarray}{rl}
             \Delta_{F}(x)  \, = \,\left(2\pi \right)^{-4}\int d^{4} q \frac{\exp\left[ i q \cdot x\right] }{m^{2} \, + \, q^{2} - i\varepsilon} \ .
    \label{FeynmanPropagator}
\end{IEEEeqnarray}
Making use of $ \left( \square - m^{2}\right)\Delta_{F}(x)  \, = \, -\delta^{4}(x) $, we write
\begin{IEEEeqnarray}{ll}
  \delta^{2}(\vartheta_{_{1}\mp})\delta^{2}  (\vartheta_{_{2}\pm}) &\tilde{\Delta}^{\pm \mp}\left( x_{_{1}},\vartheta_{_{1}},x_{_{2}},\vartheta_{_{2}}\right)   \nonumber\\
  &  \, = \,\delta^{2}(\vartheta_{_{1}\mp})\delta^{2}  (\vartheta_{_{2}\pm}) \,  \Delta_{F}\left( x^{\pm}_{_{12}}\right)  \, - \, 4  \,\delta^{4}(\vartheta_{_{1}})\,\delta^{4}  (\vartheta_{_{2}})\delta^{4}(x_{_{1}}-x_{_{2}}) \ . 
    \label{MutationRel_Causal_Fields_plusminus_plusminus}
\end{IEEEeqnarray}
The term $ \, +\, 4 i\,\delta^{4}(\vartheta_{_{1}})\delta^{4}  (\vartheta_{_{2}})\delta^{4}(x_{_{1}}-x_{_{2}}) $ is Lorentz but not supersymmetric invariant. Since this expression is local in superspace, the noncovariant part of the superpropagator induces noncovariant terms in the interactions. For the case of general superpotentials of arbitrary superfields, in order to gain some insight on their form, we recall that, although the step function  is translational and Lorentz invariant (except at spacelike separations  where to achieve Lorentz invariance commutators must vanish), it is  not supersymmetric invariant. $ \omega $ would be supersymmetric invariant if  it were evaluated at   $   x^{\pm\,0}_{_{12}} $ or even at  $  x^{0}_{_{12}}-\vartheta_{_{1}}^{\intercal}\epsilon\gamma_{5}\gamma^{0}\vartheta_{_{2}} $. Keeping in mind that the $ \Delta_{+} $ functions in \eqref{NonCorrectedSuperpropagator} are evaluated at  $   x^{\pm\,0}_{_{12}} $, we write
\begin{IEEEeqnarray}{rl}
              \omega\left( x^{0}_{_{12}}\right) \, = \,  \omega\left(  x^{\pm\,0}_{_{12}} \right) \, + \, \varsigma_{\pm} \left( z_{_{1}},z_{_{2}}\right) , \quad z=(x,\vartheta) \ ,
      \label{Averaged_StepFunction_AsSumPlusMinus}
  \end{IEEEeqnarray}
with  
  $ \varsigma_{\pm }  \left( z_{_{1}},z_{_{2}}\right) $   given by  the negative of the  next-to-zero-order fermionic expansion coefficients  in $ \omega\left(  x^{\pm\,0}_{_{12}} \right)$. The  second order  of the unitary operator in expansion \eqref{DysonRelation}  is given by
  \begin{IEEEeqnarray}{rl}
                U^{(2)}  & \, = \,{{(-i)^{2}}}\int d^{8}z_{_{1}}d^{8}z_{_{2}}\, \omega\left( x^{0}_{_{12}}\right) \mathcal{V}(z_{_{1}}) \mathcal{V}(z_{_{2}})   \ , 
    \label{SecondOrderDysonOperator}
\end{IEEEeqnarray} 
and for superpotentials can be written as
\begin{IEEEeqnarray}{rl}
                U^{(2)}                 
                   & \, = \,{{(-i)^{2}}}\int d^{8}z_{_{1}}d^{8}z_{_{2}}\,       \left[ \omega\left( x^{0}_{_{12}}\right) \mathcal{V}_{\pm}(z_{_{1}})\mathcal{V}^{*}_{\mp }(z_{_{2}})   \, + \,           \omega\left( x^{0}_{_{12}}\right)\mathcal{V}^{*}_{\mp }(z_{_{1}})\mathcal{V}_{\pm}(z_{_{1}})   \, + \, \dots\right]    \nonumber \\   
              & \, = \,    U_{\text{i}}^{(2)}   \, + \, U_{\text{n.i}}^{(2)}   \, + \, \dots \ ,
    \label{SecondOrderDysonOperatorPlusDotsTerms}
\end{IEEEeqnarray} 
with the  super Poincar\'e covariant  term 
\begin{IEEEeqnarray}{rl}
               U_{\text{i}}^{(2)}    & \, = \,  {{(-i)^{2}}}\int d^{8}z_{_{1}}d^{8}z_{_{2}}\,\left(   \omega\left(  x^{\pm\,0}_{_{12}} \right)   \mathcal{V}_{\pm}(z_{_{1}}) \mathcal{V}^{*}_{\mp}(z_{_{2}})   \, + \,   \omega\left( x^{\mp\,0}_{_{21}} \right) \mathcal{V}^{*}_{\mp}(z_{_{2}}) \mathcal{V}_{\pm}(z_{_{1}})  \right)  
       \label{UnitaryOperatorSecondOrderCovariantPart}
   \end{IEEEeqnarray} 
and  the noncovariant term 
   \begin{IEEEeqnarray}{rl}
               U_{\text{n.i}}^{(2)}    & \, = \,  {{(-i)^{2}}}\int d^{8}z_{_{1}}d^{8}z_{_{2}}\left( \,\varsigma_{\pm} \left( z_{_{1}},z_{_{2}}\right)   \mathcal{V}_{\pm}(z_{_{1}}) \mathcal{V}^{*}_{\mp}(z_{_{2}})   \, + \,  \varsigma_{\mp} \left( z_{_{2}},z_{_{1}}\right) \mathcal{V}^{*}_{\mp}(z_{_{2}}) \mathcal{V}_{\pm}(z_{_{1}})  \right)  \ . 
       \label{UnitaryOperatorSecondOrderNonCovariantPart}
   \end{IEEEeqnarray}    
Because of the fermionic delta functions in the superpotentials, we can evaluate invariant step functions at  $ \left( x^{0}_{_{12}} \, - \, 2 \vartheta_{_{1}\pm}^{\intercal}\epsilon\gamma_{5}\gamma^{0}\vartheta_{_{2}\mp}\right)  $,
allowing us to write the noncovariant part of the step functions as 
\begin{IEEEeqnarray}{rl}
                           \varsigma_{\pm}\left(z_{_{1}},z_{_{2}}  \right)            &  \,=\, 2  \vartheta_{_{1}\pm }^{\intercal}\epsilon\gamma_{5}\gamma^{0}\vartheta_{_{2}\mp }\,\delta\left( x^{0}_{_{12}}    \right) \, - \, 4\,\delta^{2}(\vartheta_{_{1}\pm }) \,\delta^{2}(\vartheta_{_{2}\mp }) \, \frac{\partial}{\partial x_{_{1}}^{0}}\,\delta\left( x_{_{12}}^{0}  \right)     \ .    
                 \label{NonInvariantStepFunctionForSuperPotentials}
             \end{IEEEeqnarray}   
We can see from this that the other terms [expressed by $ \dots$ in  \eqref{SecondOrderDysonOperatorPlusDotsTerms}] do not need to be corrected. Noting that  $    \varsigma_{-}\left(z_{_{1}},z_{_{2}}  \right)  \, = \,    -\varsigma_{+}\left(z_{_{2}},z_{_{1}}  \right)     $, we write
\begin{IEEEeqnarray}{rl}
       U_{\text{n.i}}^{(2)} & \, = \,      {{(-i)^{2}}}\int d^{8}z_{_{1}}d^{8}z_{_{2}}\,  \varsigma_{\pm}\left( z_{_{1}},z_{_{2}}\right)    \left[  \mathcal{V}_{\pm}(z_{_{1}}),  \mathcal{V}^{*}_{\mp}(z_{_{2}})   \right] \ .
    \label{UnitaryOperatorSecondOrderNonCovariantPartGen}
\end{IEEEeqnarray}  
 Using Eq. \eqref{SuperpotentialComponents}, we can integrate the fermionic variables to obtain 
\begin{IEEEeqnarray}{rl}
       U_{\text{n.i}}^{(2)} & \, = \,     +4\int d^{4}x_{_{1}}d^{4}x_{_{2}}\, \left(  i \delta\left( x^{0}_{_{12}} \right) \sum_{\alpha}\left\lbrace \left[ \Omega\left(x_{{_{1}}} \right)\right] _{\pm\alpha} ,    \left[\Omega^{*}\left(x_{{_{2}}} \right)\right]_{\pm\alpha}\right\rbrace \right.  \nonumber \\
      &\qquad  \qquad    \qquad \qquad\qquad \left.  \, + \,  \, \delta\left( x_{_{12}}^{0}  \right) \frac{\partial}{\partial x_{_{1}}^{0}}\, \left[  \mathcal{C}\left(x_{_{1}} \right)  ,   \mathcal{C}^{*}\left(x_{_{2}}\right) \right] \right)  \ .
    \label{UnitaryOperatorSecondOrderNonCovariantPartSecondTerm1}
\end{IEEEeqnarray} 
 Any (anti)commutator will generate products of fields multiplied by  $ \Delta(x) = \Delta_{+}(x) \, - \, \Delta_{+}(-x) $ functions  and derivatives.  Because of delta functions in time and $ \left( \square\Delta(x)= m^{2}\Delta(x) \right) $, the only surviving  terms in the  anticommutator(commutator)  of Eq.  \eqref{UnitaryOperatorSecondOrderNonCovariantPartSecondTerm1} are  the ones in which an odd(even) number of time derivatives act on  $  \Delta(x) $, generating four-dimensional delta functions $\delta(x^{0}_{_{12}})\frac{\partial}{\partial x_{_{1}}^{0}}\Delta\left(x_{_{1}}- x_{_{2}}\right) = -i\delta^{4}(x_{_{1}}-x_{_{2}})  $. This lets us write
\begin{IEEEeqnarray}{rl}
       U_{\text{n.i}}^{(2)} 
        & \, = \,  i  \int d^{4}x_{_{1}}d^{4}x_{_{2}}  \,\delta^{4}(x_{_{1}}-x_{_{2}}) F\left(x_{_{1}}\right)  \ ,
    \label{UnitaryOperatorSecondOrderNonCovariantPartSecondTerm2}
\end{IEEEeqnarray} 
with $  F\left(x_{_{1}}\right)    $ given explicitly by the term of the integral factor  of $ d ^{4}x_{_{1}} $ in \eqref{UnitaryOperatorSecondOrderNonCovariantPartSecondTerm1}.  Therefore, we must replace 
\begin{IEEEeqnarray}{rl}
               \mathcal{V}\left( x,\vartheta\right) \, \rightarrow  \,  \mathcal{V}\left( x,\vartheta\right)  \, + \, \delta^{4}\left(\vartheta \right) F\left( x\right) 
    \label{CorrectedPotential}
\end{IEEEeqnarray}
in order to cancel the lowest-order noncovariant term \eqref{UnitaryOperatorSecondOrderNonCovariantPartSecondTerm2}. 
It is evident that this result extends to the case of general potentials, since the effect of considering the full expansion in fermionic variables in  \eqref{Averaged_StepFunction_AsSumPlusMinus}, is to  add  further higher-order derivatives in time to Eq. \eqref{UnitaryOperatorSecondOrderNonCovariantPartSecondTerm1}. The important point is that we always have a delta function $\delta(x_{_{1}}^{0}-x_{_{2}}^{0})$, ensuring  counterterms are local in time,  the main assumption made in \eqref{PotentialAsIntegralOfTheSuperPotential}.  The  function  $   F\left(x_{_{1}}\right)     $ is in general not only supersymmetric noncovariant but also  Lorentz noncovariant. 

Equation   \eqref{CorrectedPotential}, based only on pure operator methods, has the advantage that  it gives us directly  the lowest order correction to the superpotential, but cannot discern  which  terms  are purely supersymmetric and/or Lorentz  noncovariant. More importantly, so far, we have not made  clear why  nonlocal terms cannot arise beyond the second-order Dyson operator \eqref{SecondOrderDysonOperator}.  
  We prove, at the level of the super S-matrix and on the same lines of ordinary space,   that  all of the noninvariant terms are  local. For this purpose, we introduce the function   $ P^{(L)}_{n,\tilde{n}}\left( q\right) $  for  off-shell momentum $ q $   by extending  linearly  in $ q^{0} $ the on-shell polynomial $ {P}_{n,\tilde{n}}\left(\textbf{p},p^{0} \right)$ [Eq. \eqref{LinearizedMomentumPropagator}] to the off-shell case. Therefore, the ordinary space propagator $ -i\Delta_{n,\tilde{n}}(x) $ is expressed as $ -i {P}^{(L)}_{n,\tilde{n}}   \left(-i\partial\right) \Delta_{F}(x) $ [where $ \Delta_{F} $ is \eqref{FeynmanPropagator}].  We  can always split the  function $ P_{n,\tilde{n}}^{(L)}(q) $ as the sum  of a Lorentz-covariant (polynomial in $ q^{\mu} $) part,
\begin{IEEEeqnarray}{rl}
              P_{n,\tilde{n}}^{\left( \text{off}\right) }\left( \Lambda q\right)   \, = \, \sum_{m,\tilde{m}}S_{n,m}\left(\Lambda \right) S^{*}_{\tilde{n},\tilde{m}}\left(\Lambda \right)P_{m,\tilde{m}}^{\left( \text{off}\right) }\left(  q\right)  \ ,
    \label{MomentumOffShellCovariantPropagator}
\end{IEEEeqnarray}
plus a Lorentz-noncovariant term originated at $ \left( x_{1}=x_{2}\right) $, such that when  $ q $ is  on the mass shell   $ {P}^{(L)}_{n,\tilde{n}}$  and  $ P_{n,\tilde{n}}^{\left( \text{off}\right) } $  coincide~\cite{Weinberg:1969di}. 
   By tracking first the  Lorentz noncovariant parts, we can write the general superpropagator  as
\begin{IEEEeqnarray}{rl}
       (-i){\Delta}_{n,\tilde{n}}^{\pm \mp} \left( x_{_{1}},\vartheta_{_{1}},x_{_{2}},\vartheta_{_{2}}\right)  
     &  \, = \,    (-i)\left[P^{(L)}_{n,\tilde{n}}\left( -i\partial_{_{1}}\right)  \, +\,  2 \vartheta_{_{1}}^{\intercal}\epsilon\gamma_{5}(-i\gamma^{\mu})\vartheta_{_{2}\mp} \,P^{(L)}_{\mu, n,\tilde{n}}\left( -i\partial_{_{1}}\right)\right.    \nonumber \\
      & \left.\quad  \, - \,4  m^{2}\,\delta^{2}(\vartheta_{_{1}\pm})\delta^{2}  (\vartheta_{_{2}\mp}) P^{(L)}_{n,\tilde{n}}\left( -i\partial_{_{1}}\right)  \right]  \Delta_{F}(x_{_{1}}-x_{_{2}})  \, + \, \dots \ ,     
    \label{Complete_OffShellPropagator}
\end{IEEEeqnarray}
 where  ``$ \dots $'' represents the rest of the terms in  the general fermionic expansion variables $ \vartheta_{_{1}} $ and $ \vartheta_{_{2}} $ (the explicitly shown terms are the ones that survive when we  consider the $ {\Delta}_{n,\tilde{n}}^{\pm \mp}  $ for superpotentials). The functions $ P^{(L)}(q) $ and $P_{\mu}^{(L)}(q) $ are the off-shell extensions of  the on-shell functions $ P(p) $ and $ p_{\mu}P(p) $.  We  isolate the supersymmetric and Lorentz-covariant part  $   P^{\text{(off)}}_{n,\tilde{n}}\left( -i\partial_{_{1}}\right) \Delta_{F}(x^{\pm}_{_{12}})   $  by writing \eqref{Complete_OffShellPropagator}  as ($ z_{_{i}}=(x_{_{i}},\vartheta_{_{i}}) $),
\begin{IEEEeqnarray}{rl}
             (-i) {\Delta}_{n,\tilde{n}}^{\pm \mp} \left( z_{_{1}},z_{_{2}}\right)  
     &  \, = \, (-i) P^{\text{(off)}}_{n,\tilde{n}}\left( -i\partial_{_{1}}\right) \Delta_{F}(x^{\pm}_{_{12}})  \, + \, \Upsilon_{n,\tilde{n}}\left( z_{_{1}},z_{_{2}},-i\partial_{_{1}}\right) \Delta_{F}(x_{_{1}}-x_{_{2}}) \ , 
    \label{SplitingSPoicareCov_NonCov_Spropagator}
\end{IEEEeqnarray}
with
\begin{IEEEeqnarray}{rl}
         (+i)   \Upsilon_{n,\tilde{n}}\left( z_{_{1}},z_{_{2}},-i\partial_{_{1}}\right)& \, = \, 4\delta^{2}(\vartheta_{_{1}\pm})\delta^{2}  (\vartheta_{_{2}\mp}) \left(\square  \, - \, m^{2} \right) P^{\text{(off)}}_{n,\tilde{n}}\, + \, \delta P_{n,\tilde{n}} \nonumber \\
        &    \qquad   \, +\,  2 \vartheta_{_{1}}^{\intercal}\epsilon\gamma_{5}(-i\gamma^{\mu})\vartheta_{_{2}\mp} \,\delta P_{\mu, n,\tilde{n}}\, - \,4  m^{2}\,\delta^{2}(\vartheta_{_{1}\pm})\delta^{2}  (\vartheta_{_{2}\mp}) \delta P_{n,\tilde{n}}  \, + \, \dots \  ,  \nonumber\\ 
    \label{NonCovPartOfOffShellPropagator}
\end{IEEEeqnarray}
and $ \delta P_{n,\tilde{n}} \equiv P^{{(L)}}_{n,\tilde{n}}  \, - \,P^{\text{(off)}}_{n,\tilde{n}} $ and $ \delta P_{\mu,n,\tilde{n}} \equiv P^{{(L)}}_{\mu,n,\tilde{n}}  \, - \,q_{\mu}P^{\text{(off)}}_{n,\tilde{n}} $.  The difference between  $ P^{(L)} $  and  $ P^{(\text{off})} $ must possess a factor $ q^{2}+m^{2} $ that ensures their vanishing at on-shell momentum. This factor cancels off with the denominator in \eqref{FeynmanPropagator}, giving a delta function $ \delta^{4}(x_{1}-x_{2}) $  that guarantees noncovariant terms are always local  \citep{Weinberg:1995mt}.  

It is clear that the definition of  $ P^{(L)} $ has not made $ P^{(\text{off})} $ unique, since  adding and subtracting a term $ f_{n,\tilde{n}}\left( q\right) (q^{2}+m^{2}) $ in  the covariant  and noncovariant off-shell functions, respectively, does not alter  $ P^{(L)} $, for an arbitrary  polynomial function $ f_{n,\tilde{n}}\left( q\right) $ that satisfies \eqref{MomentumOffShellCovariantPropagator}. For example, in the case of the derivative of a massive field $ \partial_{\mu}\phi $ in ordinary space, the on-shell polynomial is $ p^{\mu}p^{\nu} $, and therefore the  off-shell function is $P_{\mu\nu}^{(L)}(q)= q^{\mu}q^{\nu}  \, + \, \delta^{\nu}_{0}\delta^{\mu}_{0}\left( q^{2}+m^{2}\right) $. Any  functions of the form  $q^{\mu}q^{\nu} +\alpha \eta^{\mu\nu}\left( q^{2}+m^{2}\right)  $ and $\left( \delta^{\nu}_{0}\delta^{\mu}_{0} -\alpha \eta^{\mu\nu}\right) \left( q^{2}+m^{2}\right)  $ serve as covariant and noncovariant parts of $P_{\mu\nu}^{(L)}(q)$. In ordinary space, the choice is to take $ P^{(\text{off})} $ as the Weinberg's form \citep{Weinberg:1969di}, where the polynomial $ P_{\mu\nu}^{(L)} \, - \,P^{(\text{off})}$ has only terms that are all Lorentz noncovariant (since precisely we want to isolate those terms). In superspace, the issue is more subtle, as we explain below. 

Repeating the whole  argument that lead us to \eqref{SplitingSPoicareCov_NonCov_Spropagator}, for the pairing of $ {\Phi}_{\pm n} $ with $ \tilde{\Phi}^{*}_{\pm \tilde{n}} $, we end with a  superpropagator of the form
\begin{IEEEeqnarray}{rl}
             (-i) \Delta^{\pm\pm}_{n,\tilde{n}}\left( x_{_{1}},\vartheta_{_{1}},x_{_{2}},\vartheta_{_{2}}\right)    
              \, = \, \pm 2m(-i)\, \delta^{2}\left(\vartheta_{_{1}} \, - \, \vartheta_{_{2}} \right)_{\pm}\, \tilde{P}^{(\text{off})}_{n,\tilde{n}}   \left(-i{\partial_{_{1}}}\right) \Delta_{F}(  x^{\pm}_{_{12}})   \, + \,\dots\ 
    \label{SuperPropagator_plusminus_plusminus}
\end{IEEEeqnarray}
where ``$ \dots $'' represents the noncovariant contributions to the superpropagator. Being  completely general, we are not assuming  that  $ \tilde{P}^{(\text{off})}_{n,\tilde{n}}   \left(q\right)  $ and $ {P}^{(\text{off})}_{n,\tilde{n}}   \left(q\right)  $ coincide for the off-shell momentum, since we are only  sure that the weaker condition  holds:
\begin{IEEEeqnarray}{rl}
           {P}^{(\text{off})}_{n,\tilde{n}}   \left(p\right)    \, = \, \tilde{P}^{(\text{off})}_{n,\tilde{n}}   \left(p\right)   \, = \, {P}_{n,\tilde{n}}   \left(p\right)  , \text{  for  } p^{2}=-m^{2}\, .
    \label{WeakerFormOfPropagator}
\end{IEEEeqnarray}
Experience with canonical (or path-integral) formulations helps us to see why it is mostly the case that  $ {P}^{(\text{off})} $ has to be of the  Weinberg's form and why it is not a surprise that  $ \tilde{P}^{(\text{off})}   $ could be different from $ {P}^{(\text{off})} $. Consider general off-shell ($ \pm $)superfields $ \Phi^{\text{off}}_{\pm n}(x,\vartheta) $ of the form  
\begin{IEEEeqnarray}{rl}
                  \Phi^{\text{off}}_{\pm n}(x,\vartheta)    \, = \,  \phi_{\pm n}(x_{\pm})    \, \mp \,\sqrt{2} \vartheta_{\pm}^{\intercal}\epsilon\gamma_{5} \psi_{n}(x_{\pm})   \, \pm \,  2 \delta^{2}\left(\vartheta_{\pm}\right) \mathcal{F}_{\mp n}(x_{\pm}) \ .
    \label{OffShellSuperfield}
\end{IEEEeqnarray}
In all known formulations of supersymmetry,   $ \phi_{\pm n} $ is a {propagating} component field, while  $ \mathcal{F}_{\mp n} $ is a sum of {auxiliary} and propagating fields.  Thus, we expect that, in general,  the Green functions
$ \langle \phi_{\pm n}, \phi'^{*}_{\pm \tilde{n}} \rangle_{\text{Green}} $ and $ \langle \phi_{\pm n}, \mathcal{F}'^{*}_{\mp \tilde{n}} \rangle_{\text{Green}} $  to be different. This allows us to see that        $ {P}^{(\text{off})}_{n,\tilde{n}}   $ and $ \tilde{P}^{(\text{off})}_{n,\tilde{n}}   $   would be in general different (since to compare  the superpropagators  obtained by noncanonical and other methods it is sufficient to take one of its components in fermionic expansion) and to note that $ {P}^{(\text{off})}_{n,\tilde{n}}   $ must be of the form of Weinberg (since $ \langle \phi_{\pm n}, \phi'^{*}_{\pm \tilde{n}} \rangle_{\text{Green}} $ is made  only of propagating fields). 

The discussion of this section has revealed to us that not only the breaking of the super S-matrix Lorentz invariance but also that of its supersymmetric invariance are both due to the singularity of the commutators at the light-cone apex~\cite{Weinberg:1964cn}[see Eq. \eqref{UnitaryOperatorSecondOrderNonCovariantPartSecondTerm1}] and that by introducing noncovariant
local terms in the interaction Hamiltonian it is always possible to define a super S-matrix as fully super Poincar\'e covariant. As in the case of ordinary space, we drop the noncovariant contributions in \eqref{SplitingSPoicareCov_NonCov_Spropagator}  and \eqref{SuperPropagator_plusminus_plusminus},  assuming that the counterterms have been introduced \citep{Weinberg:1969di}.

\section{Super Feynman Rules.}\label{Sec_SuperFeynmanRules}
Having all the ingredients, now we can state the super Feynman rules.  These rules can be written in similar manner to ordinary Feynman  rules; the extra ingredient is that we have to add $ \left( \pm\right)  $  signs for every vertex formed by the superfields $ \Phi_{n+} $ and $ \Phi_{n-} $.  For a theory written as  sum of superfield polynomials $ \mathcal{H}_{\ell} $, of degree $ N_{\ell} $, the potential is
\begin{IEEEeqnarray}{rl}
            \mathcal{V}\left( x, \vartheta \right)   \, = \, \sum^{N}_{\ell}g_{\ell} \mathcal{H}_{\ell}\left( x, \vartheta \right) \ .
    \label{InteractionsAsPolynomials}
\end{IEEEeqnarray}
Now, the super Feynman rules are:~\footnote{We are following very close the form presented in Ref.  \cite{Weinberg:1964cn}.}

\noindent
{(a)} Include a factor of $ -i\,g_{\ell} $ for every vertex. \\
{(b)} For every internal line running from a  $  \left( \pm\right)  $ vertex  at $ \left(x_{_{1}},\vartheta_{_{1}} \right)  $
to a $ \left( \mp\right) $ vertex  $ \left(x_{_{2}},\vartheta_{_{2}} \right)  $, include a superpropagator:
\begin{IEEEeqnarray}{rl}                                                         
                   \left( -i \right) {P}_{n,\tilde{n}}  & \left(-i\partial _{_{1}}\right) \Delta_{F}(  x^{\pm}_{_{12}})   \ .          
    \label{Pairing_aepsilon_with_PsiEpsilonInverted}
\end{IEEEeqnarray}
{(c)} For every internal line running from a  $  \left( \pm\right)  $ vertex  at $ \left(x_{_{1}},\vartheta_{_{1}} \right)  $
 \\   to a $ \left( \pm\right) $ vertex  $ \left(x_{_{2}},\vartheta_{_{2}} \right)  $, include a superpropagator:
\begin{IEEEeqnarray}{rl}                                                         
                 \pm 2(-i)\, \delta^{2}\left(\vartheta_{_{1}} \, - \, \vartheta_{_{2}} \right)_{\pm}\, \left[m\, {P}_{n,\tilde{n}}  \right. &\left.  \left(-i\partial   _{_{1}}\right)  \Delta_{F}(  x^{\pm}_{_{12}})   \, + \,   f_{n,\tilde{n}}\left(-i\partial   _{_{1}}\right)  \delta^{4}(  x^{\pm}_{_{12}})     \right]      \ .    
    \label{Pairing_aepsilon_with_PsiEpsilon}
\end{IEEEeqnarray}
(d) For every external line corresponding to a sparticle of superspin $ j $, superspin $ z $ projection $ \sigma $ and supermomentum $ \left(p,s \right)  $, include
\begin{IEEEeqnarray}{ll}         
            \text{ $ \left( \mp \right) $-sparticle}  &\text{created at vertex $ \left( \pm \right) $}:  \nonumber \\
            & (2\pi)^{-3/2}e^{-i x\cdot p  } \, e^{   \left(   \vartheta-{2s}\right) ^{\intercal} \epsilon\gamma_{5} \, (+i\slashed{p})\,\vartheta_{\pm}}u^{*}_{n}(\textbf{p},\sigma) \ ; 
                   \label{CreatedLineSparticle_-+_+-}   \\
              \text{  $ \left( \pm \right) $-sparticle}  &\text{created at vertex $ \left( \pm \right)  $} :\nonumber \\
                  &\pm 2m  (2\pi)^{-3/2} e^{-i x_{\pm}\cdot p  }\, \delta^{2}\left[ \left(   \vartheta-s \right)_{\pm} \right] u^{*}_{n}(\textbf{p},\sigma)  \ ;
                  \label{CreatedLineSparticle_+-_+-}   \\
         \text{  $\left(  \mp \right)  $-sparticle }  &\text{destroyed at vertex $ \left( \pm \right)  $}  : \nonumber \\
          &  (2\pi)^{-3/2}e^{+i x\cdot p  }\,    e^{ -\left[ \vartheta - 2{s}\right]^{\intercal}  \epsilon\gamma_{5} \, (+i\slashed{p})\,{\vartheta}_{\pm}}u_{n}(\textbf{p},\sigma) \ ;
              \label{DestroyedLineSparticle_-+_+-}   \\
                     \text{   $\left(  \pm \right)  $-sparticle  }  &\text{destroyed at vertex $ \left( \pm \right)  $}: \nonumber \\                   
                   &   \pm 2m(2\pi)^{-3/2} e^{i\, x_{\pm}\cdot p }\,\delta^{2}\left[ \left(  s- \vartheta \right)_{\pm} \right] u_{n}(\textbf{p},\sigma)  \ ; 
                      \label{CreatedLineSparticle_+-_+-}                                                   
\end{IEEEeqnarray}
\begin{IEEEeqnarray}{ll}           
               \text{   $ \left( \mp \right) $-antisparticle }  &\text{ created at vertex $\left(  \pm \right) $}:  \nonumber \\
            & (-)^{\mathcal{B}} (2\pi)^{-3/2}e^{-i x\cdot p  }\, e^{  +\left( \vartheta- 2 s\right)^{\intercal} \epsilon\gamma_{5} \, (+i\slashed{p})\, {\vartheta}_{\pm}}\,v_{n}(\textbf{p},\sigma) \ ;
            \label{CreatedLineAntiSparticle_-+_+-}   \\
            \text{ $\left(  \pm  \right) $-antisparticle}  &\text{ created at vertex $ \left( \pm \right)  $}:  \nonumber \\
           & \pm 2m(-)^{\mathcal{B}} (2\pi)^{-3/2}e^{-i x_{\pm}\cdot p  }\, \delta^{2}\left[ \left(  \vartheta-s \right)_{\pm} \right] \,v_{n}(\textbf{p},\sigma)  \ ;
            \label{CreatedLineAntiSparticle_+-_+-}   \\                                  
              \text{  $ \left( \mp \right) $-antisparticle}  &\text{  destroyed at vertex }   \left(\pm \right) :  \nonumber \\
              & (-)^{\mathcal{B}} (2\pi)^{-3/2}e^{+i x\cdot p  }\, e^{  -\left( \vartheta -2{s}\right)  ^{\intercal} \epsilon\gamma_{5} \, (+i\slashed{p})\, { \vartheta}_{\pm}}\,v^{*}_{n}(\textbf{p},\sigma)  \ ;
            \label{DestroyedLineAntiSparticle_-+_+-}   \\               
                  \text{ $ \left( \pm \right) $-antisparticle}  &\text{  destroyed at vertex } \left( \pm\right):   \nonumber \\
                  & \pm 2m(-)^{\mathcal{B}} (2\pi)^{-3/2}e^{+i x_{\pm}\cdot p  }\,  \delta^{2}\left[ \left(  s-\vartheta \right)_{\pm} \right] \,v^{*}_{n}(\textbf{p},\sigma)  \ ;
            \label{DestroyedLineAntiSparticle_+-_+-}                                          
\end{IEEEeqnarray}
(e) Integrate all superspacetime vertex indices $ \left( x,\vartheta \right) $, etc., and sum all discrete indices
     $ n,n' $, etc. (that  come from Lorentz tensor products of the superfields in $ \mathcal{H}_{\ell} $). \\
(d) Supply minus signs that arise in theories with fermionic superfields. \\

To derive the wave superfunctions \eqref{CreatedLineSparticle_-+_+-}--\eqref{DestroyedLineAntiSparticle_+-_+-},  we have taken (anti)commutators of superfields and creation-annihilation (anti)sparticle operators.  For external legs, we can use any combination  of $ + $ or $ - $ signs, since they are related by \eqref{Massive_CreationOperators_PlusMinus_AsMinusPlusIntegrals}
and \eqref{Massive_AnnihilationOperators_PlusMinus_AsMinusPlusIntegrals}. Some remarks are pertinent:
\begin{itemize}
\item[(i)] Each vertex and each line  in the stated super Feynman rules is explicitly  super Poincar\'e covariant.  These rules  work for general supersymmetric potentials, including K\"ahler type potentials. 
\item[(ii)]Although the (presented) super  Feynman rules are formulated as superfield polynomial interactions without explicit (super)derivatives, all  number of  derivatives and all even-number of  superderivatives acting on  the superfields  are included;  any  covariant ordinary  derivative of a  $ (\pm )$ superfield is always contained in the $ (\pm) $ superfield in the tensor representation $ {\left( \mathcal{A},\mathcal{B}\right) \otimes(\frac{1}{2},\frac{1}{2})}  $  ~\citep{Weinberg:1969di}.  The superderivative product $ \mathcal{D}_{\alpha}\mathcal{D}_{\beta} $ of a $(\pm )$ superfield   is always contained in the  $(\pm )$ superfield in the representation  $ {\left( \mathcal{A},\mathcal{B}\right) \otimes(\frac{1}{2},\frac{1}{2})}  $ plus the  $(\mp )$ superfield in the representation  $ \left( \mathcal{A},\mathcal{B}\right) $,   multiplied  by a factor proportional to  $ \left\lbrace \left(I \pm \gamma_{5} \right)\epsilon\right\rbrace_{\alpha\beta} $ (see Sec. \ref{Sec_CausaleSuperfields}).
\item[(iii)] As explained at the end of Sec.  \ref{Sec_TimeOrderedSuperpropagators}, it is mostly the case that $ {P}_{n,\tilde{n}}   $ [with  the label '(off)' dropped] is of the form of Weinberg~\citep{Weinberg:1969di}. The polynomial $ f_{n,\tilde{n}}  $ is a Lorentz covariant undetermined function, that by dimensional analysis has mass dimension  equal to $ {m}{P}_{n,\tilde{n}}   $ minus  2, and this dimension is positive if superfields are chosen with canonical dimension. From this, we see that for  the case of the scalar superfield, the Weinberg's polynomial is $  {P} =1$, and therefore $ f= 0 $~\cite{Wess:1973kz}. We could have defined a new set of rules where  $ f_{n,\tilde{n}} =0 $, but it is better to leave $ f_{n,\tilde{n}}  $ general in order to easily compare the superpropagators obtained from other methods. 
\end{itemize}

\section{$  \mathit{C}$, $  \mathit{P}$, $  \mathit{T}$ and $  \mathcal{R}$ symmetries.}\label{Sec_CPTR}
To explore  the $\mathit{C}$, $  \mathit{P}$, $  \mathit{T}$  and $\mathcal{R}$ transformation properties of the superfields, we have to turn-on  the  full notation of the $ \left( {\mathcal{A},\mathcal{B}}\right)-$superfields$: \Phi_{\pm n} \rightarrow\Phi^{\mathcal{A}\mathcal{B}}_{\pm ab} $.
The transformation of annihilation and creation (anti)sparticle operators goes as follows
\begin{IEEEeqnarray}{rl}
            \mathsf{C}\,{a}_{\pm}(\textbf{p},s_{\pm},\sigma)\,\mathsf{C}^{-1}   &\, = \,\varsigma^{*}_{\pm} \varsigma\,{a}^{c}_{\pm}(\textbf{p}, \varsigma_{\pm} s_{\pm}, \sigma) \ ,\nonumber \\
               \mathsf{C}\,{a}^{c*}_{\pm}(\textbf{p},s_{\pm},\sigma)\,\mathsf{C}^{-1}   &\, = \,\varsigma_{\pm}^{c*}\varsigma^{c}\,{a}^{*}_{\pm}(\textbf{p}, \varsigma^{c}_{\pm} \, s_{\pm}, \sigma)\ ,\nonumber \\
            \mathsf{P}\,{a}_{\pm}(\textbf{p},s_{\pm},\sigma)\,\mathsf{P}^{-1}   &\, = \,\eta^{*}_{\pm}\eta\,{a}_{\mp}(-\textbf{p}, \eta_{\pm}\left( \beta s\right) _{\mp}, \sigma)\ ,\nonumber \\
               \mathsf{P}\,{a}^{c*}_{\pm}(\textbf{p},s_{\pm},\sigma)\,\mathsf{P}^{-1}   &\, = \,\eta^{c*}_{\pm}\eta^{c}\,{a}^{c*}_{\mp}(-\textbf{p}, \eta^{c}_{\pm}\left( \beta s\right) _{\mp}, \sigma)\ , \nonumber \\
            \mathsf{T}\,{a}_{\pm}(\textbf{p},s_{\pm},\sigma)\,\mathsf{T}^{-1}   &\, = \,\zeta^{*}_{\pm} \zeta(-)^{j-\sigma}\,{a}_{\pm}(-\textbf{p},\zeta_{\pm}\,\epsilon s^{*}_{\pm}, -\sigma) \ ,\nonumber \\
               \mathsf{T}\,{a}^{c*}_{\pm}(\textbf{p},s_{\pm},\sigma)\,\mathsf{T}^{-1}   &\, = \,\zeta^{c*}_{\pm} \zeta^{c} (-)^{j-\sigma}\,{a}^{c*}_{\pm}(-\textbf{p}, \zeta^{c}_{\pm}\, \epsilon s^{*}_{\pm}, - \sigma) 
    \label{ChargeConjugation_Parity_TimeReversal_OnCreationOperators}
\end{IEEEeqnarray}
where some of the phases are restricted to 
\begin{IEEEeqnarray}{rl}
            \varsigma_{+}  \, = \, \varsigma^{*}_{-}, \quad \eta_{+}  \, = \, -\eta^{*}_{-}, \quad \zeta_{+}  \, = \, -\zeta^{*}_{-}, \quad  \varsigma^{c}_{+}  \, = \, \varsigma^{c*}_{-}, \quad \eta^{c}_{+}  \, = \, -\eta^{c*}_{-}, \quad \zeta^{c}_{+}  \, = \, -\zeta^{c*}_{-}\ . \nonumber \\
    \label{RelationBetweenPhaseOfCPT}
\end{IEEEeqnarray}
The numbers that have $ \pm $ signs have to be the same for all sparticles, this in order to guarantee supersymmetric covariance (this is due to the fact that they appear in the algebra of the transformations with fermionic generators). These relations can be obtained by starting with component transformations,  then  require invariance under \eqref{Massive_NonZeroMutations_SuperCreationAnnihilationOperators} and consistency with \eqref{Massive_CreationOperators_PlusMinus_AsMinusPlusIntegrals}.  We should mention that to obtain appropriate relations for time reversal we have defined   $ \mathsf{T} s = i s^{*}\mathsf{T} $ for any fermionic number; in particular this guarantees that  $ \mathsf{T} ss'   \, = \, \left( ss'\right)^{*} \mathsf{T} $ for any pair of fermionic numbers. To perform superfield transformations, we use~\cite{Weinberg:1969di}
\begin{IEEEeqnarray}{rl}
         \left(  {u}^{\mathcal{A}\mathcal{B}}_{ab }(\textbf{p} ,\sigma)\right) ^{*}     & \, = \, \left(- \right)^{-a \, - \, b  \, -\,j} {v}^{\mathcal{B}\mathcal{A}}_{-b,-a }(\textbf{p} ,\sigma)\ , \nonumber \\
  \left( {v}^{\mathcal{A}\mathcal{B}}_{ab }(\textbf{p} ,\sigma)\right)^{*}     & \, = \,(-)^{j-a \, - \, b  }  {u}^{\mathcal{B}\mathcal{A}}_{-b,-a }(\textbf{p} ,\sigma) \ ,\nonumber \\  
        \left( {u}^{\mathcal{A}\mathcal{B}}_{ab }(\textbf{p} ,\sigma)\right)^{*}     &\, = \, \left(- \right)^{a \,+ \,  b \, + \, \sigma \, + \, \mathcal{A} \, + \, \mathcal{B}-j}  {u}^{\mathcal{A}\mathcal{B}}_{-a,-b }(-\textbf{p} ,-\sigma)\ , \nonumber \\
        \left( {v}^{\mathcal{A}\mathcal{B}}_{ab }(\textbf{p} ,\sigma)\right)^{*}     &\, = \,  \left(- \right)^{a \,+ \,  b \, + \, \sigma \, + \, \mathcal{A} \, + \, \mathcal{B}-j} {v}^{\mathcal{A}\mathcal{B}}_{-a,-b }(-\textbf{p} ,-\sigma) \ , \nonumber \\  
         {u}^{\mathcal{A}\mathcal{B}}_{ab }(-\textbf{p} ,\sigma)    &\, = \, \left(- \right)^{ \mathcal{A} \, + \, \mathcal{B}-j}  {u}^{\mathcal{B}\mathcal{A}}_{ba}(\textbf{p} ,\sigma)\ , \nonumber \\ 
          {v}^{\mathcal{A}\mathcal{B}}_{ab }(-\textbf{p} ,\sigma)    &\, = \, \left(- \right)^{ \mathcal{A} \, + \, \mathcal{B}-j}  {v}^{\mathcal{B}\mathcal{A}}_{ba}(-\textbf{p} ,\sigma) \nonumber \\       
    \label{Conjugation-Parity-Spininversion-FourierCoefficients}
\end{IEEEeqnarray}
and the properties of the exponential factor in  \eqref{CausalLeft_Right_Chiral-Sfields},
\begin{IEEEeqnarray}{rl}
            i x_{\pm} \cdot \left( \Lambda_{_{\mathcal{P}}} p\right)  & \, = \,  i\left( \Lambda_{_{\mathcal{P}}} x\right) \cdot p \, -\,\left(\varepsilon_{_{\mathcal{P}}}  \beta\vartheta\right) ^{\intercal}\epsilon\gamma_{5}\left( +i\slashed{p}\right)  \left( \varepsilon_{_{\mathcal{P}}}  \beta\vartheta \right) _{\mp} \ , \nonumber \\  
             ix_{\pm} \cdot p               & \, = \,  -\left(i\, x \cdot p \, -\,\left(\varepsilon_{_{\mathcal{C}}}\epsilon\gamma_{5}\beta \vartheta^{*}\right) ^{\intercal}\epsilon\gamma_{5}\left( +i\slashed{p}\right) \left( \varepsilon_{_{\mathcal{C}}}\epsilon\gamma_{5}\beta \vartheta^{*}\right)_{\mp}\,\right) ^{*} \ , \nonumber\\    
             \left( i x_{\pm} \cdot\left( \Lambda_{\mathcal{_{P}}}p\right)\right)^{*}            & \, = \,  i\left( \Lambda_{\mathcal{_{\mathcal{T}}}}x \right) \cdot p \, - \, (\varepsilon_{_{\mathcal{T}
              }}\epsilon\vartheta^{*})^{\intercal} \epsilon\gamma_{5} \epsilon\left( +i\slashed{p}\right)  (\varepsilon_{_{\mathcal{T}}}\epsilon \vartheta)^{*}_{\pm} \ ,
    \label{xplusminus_p_ParityTimeReversal_ChargeConjugation}
\end{IEEEeqnarray}
with $ \left( \varepsilon_{_{\mathcal{T}}}\right)^{2}  \, = \,   \left( \varepsilon_{_{\mathcal{P}}}\right)^{2}   \, = \,  - \left( \varepsilon_{_{\mathcal{C}}}\right)^{2}    \, = \,  -1$ and $ \Lambda_{_{\mathcal{T}}}  \, = \, -\Lambda_{_{\mathcal{P}}}  \, = \, \text{diag}\begin{pmatrix}
1 & 1 & 1 & -1
\end{pmatrix}  $. For a superfield transforming onto another superfield, we must impose
\begin{IEEEeqnarray}{rl}
       \eta_{+} =\eta^{c}_{+}   \, = \, \varepsilon_{_{\mathcal{P}}}, \quad       \zeta_{+} =\zeta^{c}_{+}   \, = \, \varepsilon_{_{\mathcal{T}}}, \quad \varsigma_{+} \, = \, \varsigma^{c}_{+} \, = \, \varepsilon_{_{\mathcal{C}}}  
     \label{CPT_PlusminusPhases_RelationsAfterFieldRequeriments}
 \end{IEEEeqnarray} 
 and 
 \begin{IEEEeqnarray}{rl}
       \eta^{c} =\eta(-)^{2j} , \quad \varsigma^{c} \, = \,  \varsigma, \quad \zeta  \, = \,  \zeta^{c} \ ,
     \label{CPT_ParticleAntiparticlePhases}
 \end{IEEEeqnarray} 
giving
 \begin{IEEEeqnarray}{rl}
              \mathsf{C}\, \Phi^{\mathcal{A}\mathcal{B}}_{\pm,ab}(x,\vartheta)  \,\mathsf{C}^{-1}  & \, = \,  \varsigma \left(- \right)^{2\mathcal{A}-a \, - \, b  \, -\,j} \, \Phi^{\mathcal{B}\mathcal{A}*}_{\pm,-b,-a}(x,\varepsilon_{_{\mathcal{C}}}\vartheta) \ , \nonumber \\
              \mathsf{P}    \, \Phi^{\mathcal{A}\mathcal{B}}_{\pm,ab}(x,\vartheta) \, \mathsf{P}^{-1}  &  \, = \,   \eta \left(- \right)^{ \mathcal{A} \, + \, \mathcal{B}-j}\, \Phi^{\mathcal{B}\mathcal{A}}_{\mp,ba}(\Lambda_{_{\mathcal{P}}} x, \varepsilon_{_{\mathcal{P}}} \beta\vartheta) \ ,\nonumber \\
  \mathsf{T}  \Phi^{\mathcal{A}\mathcal{B}}_{\pm,ab}(x,\vartheta)\mathsf{T}^{-1}      &\, = \,   \zeta  \left(- \right)^{a \,+ \,  b \, + \, \sigma \, + \, \mathcal{A} \, + \, \mathcal{B}-j}  \Phi^{\mathcal{A}\mathcal{B}}_{\pm,-a-b}(\Lambda_{_{\mathcal{T}}} x, \varepsilon_{_{\mathcal{T}}}\epsilon\vartheta^{*}) \ .
    \label{C_P_and_T_Superfields}
\end{IEEEeqnarray}
The combined $   \mathsf{C}  \mathsf{P}    \mathsf{T} $ transformation becomes
\begin{IEEEeqnarray}{rl}
              \left(  \mathsf{C}  \mathsf{P}    \mathsf{T}\right) \Phi^{\mathcal{A}\mathcal{B}}_{\pm,ab}(x,\vartheta)  \left(  \mathsf{C}  \mathsf{P}    \mathsf{T}\right)^{-1}   &  \, = \, \varsigma  \eta\zeta \,(-)^{2\mathcal{B}  } \Phi^{\mathcal{A}\mathcal{B}*}_{\mp,ab}(- x,  \varepsilon_{_{\mathcal{C}}}\varepsilon_{_{\mathcal{P}}} \varepsilon_{_{\mathcal{T}}} \beta\epsilon  \vartheta^{*}) \ . \nonumber \\
    \label{CPT_Superfields}
\end{IEEEeqnarray}
This last equation implies
\begin{IEEEeqnarray}{rl}
             \left(  \mathsf{C}  \mathsf{P}    \mathsf{T}\right)\mathcal{V}\left(x, \vartheta \right)  \left(  \mathsf{C}  \mathsf{P}    \mathsf{T}\right)^{-1}   \, = \, \mathcal{V}\left(-x, \varepsilon_{_{\mathcal{C}}}\varepsilon_{_{\mathcal{P}}} \varepsilon_{_{\mathcal{T}}} \beta\epsilon \vartheta^{*} \right) \ .
    \label{InteractionDensity_CPT}
\end{IEEEeqnarray}

Note that when applying $ \mathsf{T} $ to $ \mathcal{V}\left(x, \vartheta \right)  $ we pass trough $ \int d^{4}x d^{4}\vartheta $, and because $ \varepsilon_{_{\mathcal{C}}}\varepsilon_{_{\mathcal{P}}} \varepsilon_{_{\mathcal{T}}} $ is just a sign,  we can write $  \mathsf{T}d^{4}\vartheta  \, = \,\left(d^{4}\vartheta  \right)^{*}   \mathsf{T}  \, = \, d^{4}\left( \varepsilon_{_{\mathcal{C}}}\varepsilon_{_{\mathcal{P}}} \varepsilon_{_{\mathcal{T}}} \epsilon\beta\vartheta^{*}  \right)\mathsf{T}  $, giving a proof of $  \mathit{C}  \mathit{P}    \mathit{T} $ invariance for massive supersymmetric theories.

 The $ \mathcal{R} $ transformations on annihilation-creation (anti)sparticle operators are
\begin{IEEEeqnarray}{rl}           
             \mathsf{U}\left(\theta_{\mathcal{R}}\right) \,{a}_{\pm}(\textbf{p},s_{\pm},\sigma)\,\mathsf{U}\left(\theta_{\mathcal{R}}\right)^{-1}   &\, = \,e^{\left[- i(q\mp q_{0})\theta_{\mathcal{R}}\right] }\,{a}_{\pm}(\textbf{p}, e^{\left[ \mp i q_{0}\theta_{\mathcal{R}}\right] } s_{\pm}, \sigma) \ ,\nonumber \\
             \mathsf{U}\left(\theta_{\mathcal{R}}\right) \,{a}^{c*}_{\pm}(\textbf{p},s_{\pm},\sigma)\,\mathsf{U}\left(\theta_{\mathcal{R}}\right)^{-1}   &\, = e^{\left[- i(q\mp q_{0})\theta_{\mathcal{R}}\right] } {a}^{c*}_{\pm}(\textbf{p}, e^{\left[ \mp i q_{0}\theta_{\mathcal{R}}\right] } s_{\pm}, \sigma)  \ ,        
    \label{RSymmetries_CreationSparticleOperators}
\end{IEEEeqnarray} 
where  $ q_{0}$ is the same for all superparticle species. With the help of 
\begin{IEEEeqnarray}{rl}
             x_{\pm} \cdot p  \, = \, x \cdot p \, - \,\left( e^{\left[ \pm i q_{0}\theta_{\mathcal{R}}\right] } \vartheta\right) ^{\intercal}\epsilon\gamma_{5}\slashed{p} \left(e^{\left[ \mp i q_{0}\theta_{\mathcal{R}}\right] }  \vartheta\right)  _{\pm}  \ ,
    \label{ComplexChiralSpace_PropertyForRSymmetries}
\end{IEEEeqnarray}
we can write
      \begin{IEEEeqnarray}{rl}
                         \mathsf{U}\left(\theta_{\mathcal{R}}\right)\, \Phi^{\mathcal{A}\mathcal{B}}_{\pm,ab}(x,\vartheta)  \, \mathsf{U}\left(\theta_{\mathcal{R}}\right)^{-1}  & \, = \,  \exp{\left[- i(q\mp q_{0})\theta_{\mathcal{R}}\right] }\Phi^{\mathcal{A}\mathcal{B}}_{\pm,ab}(x,\mathcal{R}\vartheta) \ ,  \nonumber \\
          \label{RSymmetriesOnSuperfields}
      \end{IEEEeqnarray}
with
\begin{IEEEeqnarray}{rl}
            \mathcal{R}_{\alpha\beta}  \, = \, \begin{pmatrix}
\exp\left[ -i\theta_{\mathcal{R}}q_{0}\right]  & 0 \\ 
0 & \exp\left[ + i\theta_{\mathcal{R}}q_{0}\right]
\end{pmatrix} _{\alpha\beta}  \ .
    \label{mathCal_R}
\end{IEEEeqnarray}
In defining $ \mathcal{R} $-symmetries, we allow  $     \mathsf{U}\left(\theta_{\mathcal{R}}\right) $ to be a discrete or continuous symmetry, restricting  $ \left\lbrace \theta_{\mathcal{R}} ,q, q_{0}\right\rbrace $  to take values in a discrete set in the former case.

\section{Scalar superpotentials.}\label{Sec_ScalarSuperPotential}
In this section, we restrict ourselves to a theory of  a sparticle with zero superspin  whose interactions  are constructed with cubic polynomials of  the scalar superfield.  We calculate the lowest order correction to time-ordered products and construct a superamplitude for a sparticle-antisparticle collision.

The parity and $ \mathcal{R} $ transformations appearing in Eqs. \eqref{C_P_and_T_Superfields}  and \eqref{RSymmetriesOnSuperfields} become  
\begin{IEEEeqnarray}{rl}         
  \mathsf{P}    \, \Phi_{\pm}(x,\vartheta) \, \mathsf{P}^{-1}  &  \, = \,   \eta \Phi_{\mp}(\Lambda_{_{\mathcal{P}}} x, \varepsilon_{_{\mathcal{P}}} \beta\vartheta) \ ,\nonumber \\ 
   \mathsf{P}    \, \Phi^{*}_{\pm}(x,\vartheta) \, \mathsf{P}^{-1}  &  \, = \,   \eta^{*} \Phi^{*}_{\mp}(\Lambda_{_{\mathcal{P}}} x, \varepsilon_{_{\mathcal{P}}} \beta\vartheta) \ , \nonumber \\
               \mathsf{U}\left(\theta_{\mathcal{R}}\right)\, \Phi_{\pm}(x,\vartheta)  \, \mathsf{U}\left(\theta_{\mathcal{R}}\right)^{-1}  & \, = \,  \exp{\left[- i(q\mp q_{0})\theta_{\mathcal{R}}\right] }\Phi_{\pm}(x,\mathcal{R}\vartheta)  \ , \nonumber \\             
                         \mathsf{U}\left(\theta_{\mathcal{R}}\right)\, \Phi^{*}_{\pm}(x,\vartheta)  \, \mathsf{U}\left(\theta_{\mathcal{R}}\right)^{-1}  & \, = \,  \exp{\left[+ i(q\pm q_{0})\theta_{\mathcal{R}}\right] }\Phi^{*}_{\pm}(x,\mathcal{R}\vartheta) \ .                         
          \label{ScalarSuperFieldTransformationProperties}
      \end{IEEEeqnarray}           
For a sparticle that is its own antisparticle, the first equation in \eqref{C_P_and_T_Superfields} implies
\begin{IEEEeqnarray}{rl}
         \Phi_{\pm}(x,\vartheta) & \, = \,  \Phi^{*}_{\pm}(x,\vartheta)     \ ,
    \label{SParticleItsOwnAntiSparticle}
\end{IEEEeqnarray}     
 with $ \eta =\eta^{*} $.  For the  cubic superpotential, we have  the following  stock of possibilities  to form  interactions:
\begin{IEEEeqnarray}{rl}
            \Phi_{\pm}\Phi_{\pm}\Phi_{\pm}, \quad \Phi_{\pm}\Phi_{\pm}\Phi^{*}_{\pm}, \quad\Phi_{\pm}\Phi^{*}_{\pm}\Phi^{*}_{\pm},\quad \Phi^{*}_{\pm}\Phi^{*}_{\pm}\Phi^{*}_{\pm} \ .
    \label{MeFaltaUnTopico}
\end{IEEEeqnarray}
  Under  $ \mathcal{R} $ transformations, together with $ \delta^{2}\left( \mathcal{R}^{-1}\vartheta_{\pm} \right)  \, = \, \exp\left[ \pm 2i q_{0} \right]\delta^{2}\left(\vartheta_{\pm} \right)   $, these terms generate the following phases in the superpotential:
\begin{IEEEeqnarray}{rl}
            -3q\pm q_{0}, \quad -q\pm q_{0},\quad  +q\pm q_{0}, \quad 3q\pm q_{0} \ .
    \label{PhasesInWessZuminoModel}
\end{IEEEeqnarray}
Therefore, for  $ \mathcal{R} $-symmetric cubic superpotentials, only one term (of the four possible) survives. For a sparticle that is its own antisparticle, due to~\eqref{SParticleItsOwnAntiSparticle}, the four possibilities  shrink to one. 

Now, consider a superpotential for a sparticle with different antisparticle~\footnote{The name 'complex' superfield for such a superfield is not appropriate since superfields are always chiral.} 
\begin{IEEEeqnarray}{rl}
             \mathcal{W}_{+}\left( x, \vartheta \right)   \, = \,  \frac{g_{+}}{3!}  \left( \Phi_{+}\left( x, \vartheta \right)\right)  ^{3}  \, + \, \frac{ g_{-}}{3!} \left( \Phi^{*}_{+}\left( x, \vartheta \right) \right) ^{3}\ ,  \nonumber \\
              \mathcal{W}^{*}_{-}\left( x, \vartheta \right)   \, = \,  \frac{g^{*}_{-}}{3!} \left( \Phi_{-}\left( x, \vartheta \right)\right)  ^{3}  \, + \, \frac{g^{*}_{+}}{3!} \left( \Phi^{*}_{-}\left( x, \vartheta \right) \right) ^{3}  \ .
     \label{ScalarSuperPotentialExampleCubicSuperPotential}
 \end{IEEEeqnarray} 
 When  either $ g_{+} $ or $ g_{-} $ is zero, if $ \mathcal{R} $-charges are properly chosen, we obtain $ \mathcal{R} $-invariant superpotentials.   

From \eqref{SuperpotentialComponents} and \eqref{SuperFieldInComponents}, we can see that 
 \begin{IEEEeqnarray}{rl}
             \mathcal{C}\left(x \right)  &\, = \, \frac{g_{+}}{3!}  \left( \phi_{+}\right)^{3}  \, + \, \frac{g_{-}}{3!}\left( \phi^{*}_{-}\right)^{3} \ ,\nonumber \\
 \Omega\left(x\right)   &\, = \ -\frac{g_{+}}{2}  \left(  \phi_{+} \right) ^{2}\psi  \, + \, \frac{g_{-}}{2} \left(  \phi^{*}_{-} \right) ^{2}\left[ \epsilon\gamma_{5}\beta  \psi^{*}\right]  \nonumber \\
 \mathcal{F}\left(x \right)    &\, = \,{g_{+}} \left( -  \phi_{+} \psi ^{\intercal}    \epsilon\psi_{+}    \, + \,   m\left( \phi_{+}\right)^{2} \phi_{-}\right) \ ,\nonumber \\
 &\qquad    \, + \, {g_{-}}\left(-  \phi^{*}_{-} \psi ^{\dagger}    \epsilon\psi^{*}_{-}    \, + \,  m \left( \phi^{*}_{-}\right)^{2} \phi^{*}_{+} \right)  \ .  
    \label{ComponentsForThreeLinearTerms}
\end{IEEEeqnarray}
For this superpotential, the two lowest order correction terms  in \eqref{UnitaryOperatorSecondOrderNonCovariantPartSecondTerm1} are~\footnote{To prepare us for field theory, we ignored bilinear terms when we brought  $   \left[  \mathcal{C}\left(x_{_{1}}\right) , \mathcal{C}^{*}\left(x_{_{2}}\right)\right]   $ to the form \eqref{CommutatorCorrections}.}
\begin{IEEEeqnarray}{rl}
    i\delta\left(x^{0}_{_{12}} \right)  \sum_{\alpha}\left\lbrace \left[ \Omega\left(x_{{_{1}}} \right)\right] _{\pm\alpha} ,    \left[\Omega^{*}\left(x_{{_{2}}} \right)\right]_{\pm\alpha}\right\rbrace  & \, = \,      -2 \delta\left( {x_{_{12}}^{0}} \right) \tfrac{\partial}{\partial x^{0}_{_{1}}}    \left[  \mathcal{C}\left(x_{_{1}}\right) , \mathcal{C}^{*}\left(x_{_{2}}\right)\right]    \nonumber \\    
           &   \, = \,     
              \tfrac{1}{2} \left[ i\delta^{4}\left(x_{_{1}}-x_{_{2}}\right)  \right] \, F\left( x_{_{2}}\right)  \  , \label{CommutatorCorrections}
\end{IEEEeqnarray}
where  $  F\left( x_{_{2}}\right) $  is the function appearing in  \eqref{UnitaryOperatorSecondOrderNonCovariantPartSecondTerm2} given by
\begin{IEEEeqnarray}{rl}
             F\left(  x_{_{2}}\right)  \, = \,  \vert g_{+} \vert^{2}\left( \phi_{+}\left(x_{_{2}}\right) \right)^{2} \left( \phi^{*}_{+}\left(x_{_{2}}\right) \right)^{2}   \, + \,  \vert g_{-} \vert^{2}\left( \phi_{-}\left(x_{_{2}}\right) \right)^{2} \left( \phi^{*}_{-}\left(x_{_{2}}\right) \right)^{2}  \ .
     \label{FirstCorrectionToThePotentials}
 \end{IEEEeqnarray}
 
The covariant spacetime potential 
\begin{IEEEeqnarray}{rl}
           -i V\left(x\right)    &  \, = \,   \mathcal{F}\left(x \right)  \, - \, \, \mathcal{F}\left(x \right) ^{*}
    \label{NotCorrectedCovariantPotential}
\end{IEEEeqnarray}
acquires the form
\begin{IEEEeqnarray}{rl}
           -i V\left(x\right)    
             & \, = \,  {g_{+}} \left( -  \phi_{+}\psi ^{\intercal}    \epsilon\psi_{+}   + \, m \ \left( \phi_{+}\right)^{2} \phi_{-}\right)  \, + \, 
  {g_{-}}\left( - \phi_{-} \psi ^{\intercal}    \epsilon\psi_{-}    \, - \,   m\left( \phi_{-}\right)^{2} \phi_{+}\right) \nonumber \\
 & \qquad  \, + \, {g^{*}_{-}}\left(- \phi^{*}_{-} \psi ^{\dagger}    \epsilon\psi^{*}_{-}    \, + \, m  \left( \phi^{*}_{-}\right)^{2} \phi^{*}_{+} \right)  \, + \, {g^{*}_{+}}\left(-\phi^{*}_{+} \psi ^{\dagger}    \epsilon\psi^{*}_{+}    \, -\, m \left( \phi^{*}_{+}\right)^{2} \phi^{*}_{-} \right) \ .\nonumber \\
    \label{NotCorrectedCovariantPotential2}
\end{IEEEeqnarray}
 Finally, after integrating the fermionic variables in  \eqref{CorrectedPotential}, the resulting corrected spacetime potential is  
\begin{IEEEeqnarray}{rl}
             -\mathcal{H}_{\text{int}} \left(x\right)  & \, = \,  - F\left( x\right)   \, - \,V\left(x\right)    \nonumber \\ 
   &\, = \,    -i{g_{+}} \left( -  \phi_{+}\psi ^{\intercal}    \epsilon\psi_{+}   + \, m \ \left( \phi_{+}\right)^{2} \phi_{-}\right)  \, - \, i{g_{-}}\left( - \phi_{-} \psi ^{\intercal}    \epsilon\psi_{-}    \, - \,   m\left( \phi_{-}\right)^{2} \phi_{+}\right) \nonumber \\
 & \qquad  \, - \, i{g^{*}_{-}}\left(- \phi^{*}_{-}  \psi  ^{\dagger}    \epsilon\psi^{*} _{-}    \, + \, m  \left( \phi^{*}_{-} \right)^{2} \phi ^{*}_{+} \right)   \, - \, i{g^{*}_{+}}\left(-\phi^{*}_{+}  \psi  ^{\dagger}    \epsilon\psi^{*}  _{+}    \, -\, m \left( \phi^{*}_{+}\right)^{2} \phi^{*}_{-}   \right)  \nonumber \\
   &\qquad -\left( \vert g_{+} \vert^{2}\left( \phi_{+}\right) ^{2}\left( \phi^{*}_{+}  \right) ^{2}  \, + \, \vert g_{-}  \vert^{2}\left( \phi_{-}\right) ^{2}\left( \phi^{*}_{-}  \right) ^{2}\right) \ .
    \label{CorrectedLagrangianCubicScalarSuperfield}
\end{IEEEeqnarray}

For the case when a particle is its own antiparticle, the component fields satisfy 
\begin{IEEEeqnarray}{rl}
              \phi  \, = \, \phi_{+} \, = \, \phi^{*}_{-}, \quad \epsilon\gamma_{5}\beta\psi  \, = \, -\psi^{*}\ .
    \label{ComponentFieldsScalarAntisParticleSameAsSparticle}
\end{IEEEeqnarray}
The most general  (corrected) spacetime cubic potential for this case is
\begin{IEEEeqnarray}{rl}
               -\mathcal{H}'_{\text{int}} \left(x\right)  & \, = \,   -i{g} \left( + \phi\bar{\psi} \psi_{+}   + \, m  \left( \phi\right)^{2} \phi^{*}\right)     \, + \, i{g^{*}}\left(\phi^{*}  \bar{\psi} \psi_{-}     \, + \,  m \left( \phi^{*}\right)^{2} \phi   \right)  \, - \,\vert g \vert^{2}  \left( \phi\right) ^{2}\left( \phi^{*}  \right) ^{2} \ . \nonumber \\ 
    \label{LagrangianCubicScalarSameSparticleAndAntiSparticle}
\end{IEEEeqnarray}
Making  $ ig    \, = \, \sqrt{2}\lambda\, e^{+i\alpha} $ and $  {\sqrt{2}}\, \phi    \, = \, {e^{-i\alpha}}\left( A + iB\right) $, this last equation can be written as
\begin{IEEEeqnarray}{rl}
             -\mathcal{H}'_{\text{int}} \left(x\right)   &\, = \,     -  \lambda \, A \left( \bar{\psi} \psi\right)   \, - \,i\lambda \, B \left( \bar{\psi} \gamma_{5}\psi\right)   \, - \, m  \lambda\, A\left( A^{2}  \, + \, B^{2}\right)   \, - \, \tfrac{\lambda^{2}}{2}\left( A^{2}  \, + \, B^{2} \right) ^{2}   \nonumber\\
    \label{LagrangianCubicScalarSameSparticleAndAntiSparticleOriginalForm}
\end{IEEEeqnarray}
which is the interaction Lagrangian of the Wess-Zumino model\cite{Wess:1973kz}. Thus, Eq.~\eqref{CorrectedLagrangianCubicScalarSuperfield} generalizes to the case where a sparticle is different from its antisparticle and where possibly parity and $ \mathcal{R} $ symmetries are not conserved.

 We now are ready to compute a superamplitude of  a  sparticle-antisparticle process for either $ g_{+} $ or $ g_{-} $  zero  in \eqref{ScalarSuperPotentialExampleCubicSuperPotential}.
\begin{center}
\begin{tikzpicture}[line width=1.5 pt, scale=1.3]	
\begin{scope}[shift={(4,0)}]
	\begin{scope}[rotate=90]
			\draw[fermionbar] (-140:1)--(0,0);
			\draw[fermionbar] (140:1)--(0,0);
			\draw[fermion] (0:1.5)--(0,0);
			\node at (-140:1.2) {$2$};
			\node at (140:1.2) {$1^{c}$};
			\node at (.1,.2) {$\pm$};
			\node at (1.4,.2) {$\mp$};			
		\begin{scope}[shift={(1.5,0)}]
			\draw[fermion] (-40:1)--(0,0);
			\draw[fermion] (40:1)--(0,0);
			\node at (-40:1.2) {$2^{c}$};
			\node at (40:1.2) {$1$};	
		\end{scope}		
	\end{scope}		
\end{scope}
\end{tikzpicture}\\
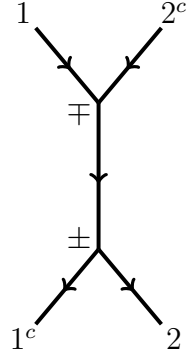
 \, Figure 1. Lowest order superdiagram for a sparticle-antisparticle collision.
 \label{Figure 1.}
\end{center}

To lowest order, there is only one superdiagram for a sparticle-antisparticle collision (Figure 1). 
For the external legs, we choose left or right fermionic 4-spinors as follows:
\begin{IEEEeqnarray}{rl}
            1\rightarrow\pm, \quad 1^{c}\rightarrow\mp,  \quad 2\rightarrow\mp, \quad 2^{c}\rightarrow\pm \ . 
    \label{adfadsf}
\end{IEEEeqnarray}
The upper (lower) signs correspond to the case $ g_{-} = 0$ ($ g_{+} = 0$). 
 After integrating out configuration superspacetime variables, we are left with
\begin{IEEEeqnarray}{rl}
S_{g_{\mp}}\left(\textbf{p}_{_{1}},s_{_{1}\pm},\textbf{p}^{c}_{_{1}}, s^{c}_{_{1}\mp},\textbf{p}_{_{2}},s_{_{2}\mp},\textbf{p}^{c}_{_{2}}, \right. & \left.  s^{c}_{_{2}\pm}\right)   \, = \,  \nonumber \\
         \left( -4i\right)  &\vert g_{\mp}\vert^{2}  f \left( \textbf{p}_{_{1}},\textbf{p}^{c}_{_{1}},\textbf{p}_{_{2}},\textbf{p}^{c}_{_{2}}\right) \times  \frac{  \left( p^{c}_{_{1}} \, - \, p_{_{2}}\right) ^{2}}{m^{2}+\left( p^{c}_{_{1}} \, - \, p_{_{2}} \right) ^{2}}\nonumber \\
          \nonumber \\
        \times\exp & { \left\lbrace  -2i\left( \slashed{p}^{c}_{_{2}} s_{_{2}}^{c} \, - \, \slashed{p}_{_{1}} {s}_{_{1}}\right)^{\intercal} \epsilon\gamma_{5}\tfrac{  \left( \slashed{p}^{c}_{_{1}} \, - \,\slashed{p}_{_{2}}\right)}{\left( p^{c}_{_{1}} \, - \, p_{_{2}}\right) ^{2} } \left(\slashed{p}_{_{2}} s_{_{2}} \, - \, \slashed{p}^{c}_{_{1}} {s}^{c}_{_{1}} \right)_{\pm}\right\rbrace  }\ ,  \nonumber \\           
    \label{LowestOrderSuperDiagrammCubicSuperpotential}
\end{IEEEeqnarray}
where
\begin{IEEEeqnarray}{rl}
            f \left( \textbf{p}_{_{1}},\textbf{p}^{c}_{_{1}},\textbf{p}_{_{2}},\textbf{p}^{c}_{_{2}},\right)   \, = \, (2\pi)^{-2} \left[ 16\left( p_{_{1}}\right)^{0} \left(   p_{_{1}}^{c}\right)^{0} \left(   p_{_{2}} \right)^{0} \left(   p_{_{2}}^{c} \right)^{0} \right]^{-1/2} \,\delta^{4}\left(p_{_{1}} \, + \, p^{c}_{_{1}} \, - \, p^{c}_{_{2}}  \, - \, p_{_{2}} \right) \ . \nonumber\\
    \label{f_term_InSuperamplitudeTwoSparticleCubicSuperPotential}
\end{IEEEeqnarray}

To calculate the particle-antiparticle scattering-amplitude for particles that are created by the~\footnote{$ a^{*}_{+} $  and $ a^{*c}_{-} $  for $ g_{-}=0 $  and $ a^{*}_{-} $  and $ a^{*c}_{+} $ for $ g_{+}=0 $.} $ a^{*}_{\pm}\left(\mathbf{p} \right) $  and $ a^{c*}_{\mp}\left(\mathbf{p} \right) $,  we take  $ s_{_{1}\pm} \, = \,  s^{c}_{_{1}\mp} \, = \, s_{_{2}\mp} \, = \,  s^{c}_{_{2}\pm} \, = \, 0 $ and the exponential factor  in \eqref{LowestOrderSuperDiagrammCubicSuperpotential} vanishes. Then, since
\begin{IEEEeqnarray}{rl}
             \frac{    \left( p^{c}_{_{1}} \, - \, p_{_{2}}\right) ^{2}}{m^{2}+\left( p^{c}_{_{1}} \, - \, p_{_{2}} \right) ^{2}}  \, = \,   1\, - \,  \frac{ m^{2}}{m^{2}+\left( p^{c}_{_{1}} \, - \, p_{_{2}} \right)^{2}} \ ,
    \label{SumOfTwoDiagramsCase}
\end{IEEEeqnarray}
 the zero component of the superamplitude  is giving  us the sum of two Feynman diagrams. These diagrams correspond to the interaction  terms [present in \eqref{CorrectedLagrangianCubicScalarSuperfield}]:
\begin{IEEEeqnarray}{rl}
              (\mp i m )\, g_{\mp}\left(\phi_{\mp}\right)^{2}\phi_{\pm}  \, + \, \text{H.c} \, + \,  \vert g_{\mp} \vert^{2}\left( \phi_{\mp}\right) ^{2}\left( \phi^{*}_{\mp}  \right) ^{2}  \ .
    \label{TwoTermsSFeynmanOrderZero}
\end{IEEEeqnarray}

 The  particle-antiparticle scattering with three particles and three antiparticles gives us a total of $ 3^{4} $ initial-final state combinations~\footnote{Some of them are zero, for example all odd fermionic expansions in \eqref{LowestOrderSuperDiagrammCubicSuperpotential}.}.  Therefore, Eq. \eqref{LowestOrderSuperDiagrammCubicSuperpotential} represents a very economical expression for the set of all processes of these particles at order $ \vert g_{\mp} \vert^{2} $.

\section{Conclusions and outlook.}\label{Sec_Conclusions} 

In this paper, we obtain perturbative scattering superamplitudes as super Feynman diagrams for sparticles and antisparticles  that carry any superspin. We accomplish this by introducing interactions out of  superfields $ \Phi_{+ n} $, $ \Phi_{- n} $, and their  adjoints, in any representation $ \left( \mathcal{A},\mathcal{B}\right)  $ of the Lorentz group. These superfields possess component fields $ \phi_{+ n} $, $ \phi_{-n} $ in the representation  $ \left( \mathcal{A},\mathcal{B}\right)  $  and $ \psi_{n} $ in  the representation  $ \left[ \left(\frac{1}{2},0 \right) \oplus  \left(0 , \frac{1}{2}\right)  \right]\otimes \left( \mathcal{A},\mathcal{B}\right)   $.

It is striking that for scalar superfields, as we know from canonical and path integral formulations, the lowest-order
correction to time-ordered products seems to be necessary and sufficient to guarantee supersymmetric invariance at all
orders, suggesting that perturbatively some sort of domino  effect mechanism is occurring: lowest-order corrections
introduced at first order in Dyson series are canceling noncovariant terms in second order, and these corrections
then generate second-order terms that seem to be canceling the noncovariant terms arising at third order, and so on.
Since fermionic expansion coefficients of superamplitudes are picking up external lines, to any order in coupling
constants, these coefficients are giving the sum of all possible diagrams originated at that order.

Pertubartively, most broken supersymmetric theories preserve the particle number of exact supersymmetric
theories. Thus, the formalism presented in this work can in principle be extended to compute superamplitudes in
phases of the theory where nondegeneracy of the supermultiplet masses is unimportant. This can be done by
extending the super Feynman rules to include symmetry breaking terms that originate as local couplings constants in
the fermionic variables.

Generalizations to the $ \mathcal{N} $-extended supersymmetry case seem straightforward, since the obtained creation-annihilation superparticle operators, presented in Sec. \ref{Sec_SuperStates},  admit a recursive procedure: creation-annihilation superparticle operators in $ \mathcal{N} $-extended momentum superspace can be defined in terms of the creation-annihilation superparticle operators in $ (\mathcal{N}-1) $-extended momentum superspace.

The proposal may find applications beyond those of higher superspin theories for example by extending results in operator-based formulations of quantum field theory to the superspace case. The obtention of multiparticle superstates $\left|\mathcal{N}\right. \rangle $ that transform fully covariant under arbitrary super Poincar\'e transformations makes it possible to express the general matrix element $ \langle  \mathcal{M}\left| \mathcal{O}(z_{_{1}},\dots, z_{_{n}})\right|\mathcal{N}\rangle $ for  superspace operators $ \mathcal{O}$ (created with Heisenberg  superfields evaluated at $ (z_{_{1}},\dots, z_{_{n}}) $ and possibly time ordered) as matrix elements at arbitrary shifted values $ z_{_{1}}-z,\dots,  z_{_{n}}-z  $.   This shifting is used in intermediary matrix elements that are present in some operator-based works, such as the spectral representations~\cite{Kallen:1952zz,Lehmann:1954}, the operator product expansion (OPE)~\cite{Wilson:1969zs}, and spontaneously global symmetries~\cite{Goldstone:1962es}. So far, superspace extensions to these results have been presented only in the context of functional-based approaches (the supersymmetric Kallen–Lehmann representation and the OPE for the scalar superfield are offered in  Refs.~\cite{Leroy:1986ve,Constantinescu:2003vn}). Also, it could be useful to write fully supersymmetric covariant results that are usually present in component form, such as the kinematical constraints in supergravity~\cite{Grisaru:1976vm} and the tree QCD amplitudes from supersymmetric scattering amplitudes~\cite{Dixon:1996wi}.   In addition, midway between Lagrangian and pure S-matrix formulations, the super Feynman rules for arbitrary massless superparticles should be straightforward~ \cite{Weinberg:1964ev} (but it will be instructive to compare it with the zero mass limit of our results), superspace investigations for the higher-dimensional theories~\cite{Weinberg:1984vb}, and scale and conformal invariant field theories~\cite{Chan:1973iq,Weinberg:2012cd} seem also very well suited. To obtain general super wave functions for supersymmetric gauge theories and gravitation will be more challenging, but extensions along the lines of Refs. \cite{Weinberg:1965rz,Weinberg:1965nx,Grisaru:1976vm} seem feasible (from which evidence of new soft theorems  and relations with new Ward identities have recently been found \cite{Cachazo:2014fwa,He:2014laa}).

\begin{acknowledgments}
  I would like to thank Myriam Mondrag\'on  and Carlos Vaquera  for useful discussions and for corrections to the manuscript. I  acknowledge support from the Mexican grants: PAPIIT IN113712 and CONACyT-132059.
  \end{acknowledgments}
 \appendix
\section{Notation and conventions. }\label{Sec_ConventionsNotation}
We use repeatedly identities of Dirac matrices and  fermionic 4-spinor variables. Since these relations are  standard, we limit ourselves to present the notation and conventions employed in the paper.  We represent Dirac and Lorentz indices by $ \alpha,\alpha',\beta,\beta' $, etc., and  $ \mu,\nu,\mu',\nu' $, etc., respectively. We take the Lorentz metric as $ \eta_{\mu\nu}  \, = \, \text{diag}\begin{pmatrix}
1 & 1 & 1 & -1
\end{pmatrix}  $. The Dirac Representation $ D(\Lambda) $ is generated by
	\begin{equation}
      D\left[ \Lambda\right]   \, = \, \exp\left[\tfrac{i}{2} w_{\mu\nu}\mathcal{J}^{\mu \nu} \right], \quad     \mathcal{J}^{\mu \nu} \, = \, \tfrac{-i}{4}[\gamma^{\mu}, \gamma^{\nu}] \ , 
         \label{Appendix_AlgebraMatricesDirac}
	\end{equation}
where  the anticommutator of  $ \gamma $-matrices is taken positive:  $  \left\lbrace  \gamma^{\mu} ,  \gamma^{\nu}\right\rbrace   \, = \, 2\eta^{\mu\nu} $. We stick to the representation
\begin{equation}
         \gamma^{0}\, = \,-i\begin{pmatrix}
0 & I \\ 
I & 0
\end{pmatrix}  \, = \, -i\beta, \quad {\gamma}_{i}\, = \,-i  \begin{pmatrix}
0 & \sigma_{i} \\ 
-\sigma_{i} & 0
\end{pmatrix}\ .
         \label{Appendix_GammaWeylBasis}
	\end{equation}
Also, we use
\begin{IEEEeqnarray}{rl}
     \gamma_{5}  \, = \, \begin{pmatrix}
I & 0 \\ 
0 & -I
\end{pmatrix} 
, \quad         \epsilon  \, = \, \begin{pmatrix}
 e & 0 \\ 
 0 & e
 \end{pmatrix} , \quad e  \, = \, \begin{pmatrix}
 0 & 1 \\ 
 -1 & 0
 \end{pmatrix} ,
    \label{Appendix_Transposing}
\end{IEEEeqnarray}
that together with $ \beta $ satisfy
 \begin{IEEEeqnarray}{rl}
               \beta \gamma^{\mu}   \, = \,  -\gamma^{\mu\dagger}\beta, \quad   \epsilon\gamma_{5} \gamma^{\mu} \, = \, -\gamma^{\mu \intercal}\epsilon\gamma_{5}\ .
     \label{Appendix_Transpos_ConjugateOfDiracMatrices}
 \end{IEEEeqnarray}	
For the standard transformation $ p  = L(p)k $, we take $ k =\begin{pmatrix}
0 & 0 & 0 & m
\end{pmatrix}  $ as standard vector. 

 For any 4-spinor $ v $, its left projection is  written as $ v_{+}=\frac{1}{2}(I+\gamma_{5})v $ and its right projection as $ v_{-}=\frac{1}{2}(I-\gamma_{5})v $.  Useful identities for fermionic 4-spinors are
\begin{IEEEeqnarray}{rl}
               (s_{\pm})(s_{\pm})^{\intercal}   &\, = \,\tfrac{1}{2}\left[ \epsilon \left( I \pm \gamma_{5}\right) \right] \delta^{2}(s_{\pm})  \ , \nonumber \\ 
                (s_{\pm})(\epsilon\gamma_{5} s)_{\mp}^{\intercal}   &\, = \,\tfrac{1}{4}\left(  s^{\intercal}\epsilon\gamma_{5}\gamma_{\mu}s_{\pm} \right) \left[  I \, \pm \, \gamma_{5}\right]\gamma^{\mu}     \ , \nonumber \\
                 s^{\intercal}\epsilon\gamma_{5}\gamma_{\mu}s_{\pm} &  \, = \,-s^{\intercal}\epsilon\gamma_{5}\gamma_{\mu}s_{\mp}   \ , \nonumber \\
                      \left( s^{\intercal}\epsilon\gamma_{5}\gamma_{\mu}s_{\pm}\right) ^{*} &  \, = \,\left( \epsilon\gamma_{5}\beta s^{*}\right) ^{\intercal}\epsilon\gamma_{5}\gamma_{\mu}\left( \epsilon\gamma_{5}\beta s^{*}\right)_{\pm} \ ,
    \label{zetaPlusMinus_quad}
\end{IEEEeqnarray}  
 where  $ \delta ^{2}(s) $ is defined by
 \begin{IEEEeqnarray}{rl}
            \delta ^{2}(s)  &\, \equiv \, \tfrac{1}{2}  {s}^{\intercal} \epsilon s , \quad 
               \left[   \delta^{2}(s)\right]^{*}    \, = \,    - \delta^{2}(s^{*})   \ .
    \label{Appendix_Equality_zetaPlusMinusAlpha_zetaPlusMinusBeta}
\end{IEEEeqnarray}

A 4-spinor satisfies the Majorana condition if
\begin{IEEEeqnarray}{rl}
            s  \, = \, \epsilon\gamma_{5}\beta s^{*} \ .
    \label{Appendix_MajoranaCondition}
\end{IEEEeqnarray}

\section{Fermionic Integrals.}\label{Sec_BerezinIntegrals}
 Given a set of fermionic variables $ \zeta_{1}\dots \zeta_{N} $, the Berizinian integral is defined to act from the left,
\begin{IEEEeqnarray}{rl}
            \int d\zeta_{N'}\dots d\zeta_{2}d\zeta_{1}\,\left\lbrace \zeta_{1}\zeta_{2}\dots \zeta_{N'} A\right\rbrace   \, = \, A ,\quad   N'\leq N \ .
    \label{Appendix_DefinitionLeftBerezinIntegral}
\end{IEEEeqnarray}
The lowest dimension (nontrivial) integral with this set of fermionic variables is the line integral,
\begin{IEEEeqnarray}{rl}
               \sum_{ij}^{n}\int d{\zeta}_{i}^{\intercal}\zeta_{j}C_{ij} \, = \,  \text{Tr}C  \, = \,  \sum_{ij}^{n}\int d{(D\zeta)}_{i}^{\intercal} \left( D\zeta \right)_{j}C_{ij}\ ,
     \label{Appendix}
 \end{IEEEeqnarray} 
 where  $ D_{ij} $ is an invertible bosonic  matrix; since $ \text{Tr}C   \, = \, \text{Tr}D^{-1}C D $ we have $ d{(D \zeta)^{\intercal}}  \, = \,d{\zeta^{\intercal}}   D^{-1} $. This  holds for any surface Berezinian integral:
\begin{IEEEeqnarray}{rl}
            d{(D \zeta)}_{{1}}d{(D \zeta)}_{{2}}\dots d{(D \zeta)}_{N'}  \, = \, 
            \left[ \left( D^{-1} \right)^{\intercal} d \zeta\right] _{1}\left[ \left( D^{-1} \right)^{\intercal} d \zeta\right] _{2}\dots\left[ \left( D^{-1} \right)^{\intercal} d \zeta\right] _{N'} \ . \nonumber \\
    \label{Appendix_SurfaceElementTransformation}
\end{IEEEeqnarray} 
The right side of the  complex conjugate of \eqref{Appendix_DefinitionLeftBerezinIntegral}
  is  $ A^{*} $. If we allow  conjugation to enter in the integral as $\left(  \zeta_{1}\zeta_{2}\dots\zeta_{N} \,  \right)^{*}  $, the net effect in the integral is  
\begin{IEEEeqnarray}{l}
               \left( \int d\zeta_{N'}\dots d\zeta_{2}d\zeta_{1}\,\left\lbrace \zeta_{1}\zeta_{2}\dots \zeta_{N'} A\right\rbrace \right)^{*}   \, = \,  \int   \left( d\zeta_{N'}\dots d\zeta_{2}d\zeta_{1}\right) ^{*}\, \left(\zeta_{1}\zeta_{2}\dots\zeta_{N'}  \right)^{*}  A^{*} \ .\nonumber \\
      \label{Appendix_ConjugationOfFermionicIntegral}
  \end{IEEEeqnarray}  
For  fermionic 4-spinors, two dimensional and four dimensional fermionic differentials are defined by
\begin{IEEEeqnarray}{rl}
            d^{2}s_{\pm}  \, \equiv \, -\tfrac{1}{2} ds^{\intercal}_{\pm} \epsilon ds_{\pm}, \quad d^{4}s  \, \equiv \,  d^{2}s_{+}   d^{2}s_{-} \ .  
    \label{Appendix_TwoFourDiffFermiSpinor}
\end{IEEEeqnarray}
They give
\begin{IEEEeqnarray}{rl}
             \int d^{2}s_{\pm} \delta^{2}\left( s_{\pm}\right) = \int d^{4}s \,\delta^{4}\left( s\right)  \, = \, 1 \ ,
    \label{Appendix_TwoFourDiffFermiSpinor_FermionicWithDeltaFunctions}
\end{IEEEeqnarray}
where  $ \delta^{4}\left( s\right) =  \delta^{2}\left( s_{+}\right) \delta^{2}\left( s_{-}\right)$. Under conjugation,
\begin{IEEEeqnarray}{rl}
            \left( d^{2}s_{\pm}\right)^{*}   \, = \,  -d^{2} s^{*} _{\pm}\, \quad \left( d^{4}s\right)^{*}   \, = \,d^{4} s^{*} \ .
    \label{Appendix_Conjugation_TwoFourDiffFermiSpinor}
\end{IEEEeqnarray}
From \eqref{Appendix_SurfaceElementTransformation}, we have
\begin{IEEEeqnarray}{rl}
                d^{4} s^{*}  \, = \, d^{4} \left( \epsilon s^{*}\right)  \, = \,d^{4} \left( \gamma_{5} s^{*}\right)   \, = \, d^{4} \left(\beta s^{*}\right)\, = \, d^{4} \left( \epsilon\gamma \beta s^{*}\right)  \  .   
       \label{Appendix_Invariance_FourDiffConjugated}
   \end{IEEEeqnarray}   
For an arbitrary operator density $  \mathcal{K}\left(s\right)  $ that appears as
\begin{IEEEeqnarray}{rl}
            \int d^{4}s \,\mathcal{K}\left(s\right),
    \label{Appendix_GeneralDensity}
\end{IEEEeqnarray}
 due  to \eqref{Appendix_ConjugationOfFermionicIntegral} and \eqref{Appendix_Invariance_FourDiffConjugated},  Hermiticity and Lorentz invariance in the higher-order fermionic expansion $ s $   of $ \mathcal{K}\left(s\right) $ can be chosen   as the requirement that
\begin{IEEEeqnarray}{rl}
              \mathcal{K}\left(s\right)  \, = \,    \left[ \mathcal{K}\left(\epsilon\gamma_{5}\beta s^{*}\right)\right]^{*} \ .
    \label{Appendix_HermiticityCondition}
\end{IEEEeqnarray}
If $ s $ satisfies the Majorana condition \eqref{Appendix_MajoranaCondition}, then Eq. \eqref{Appendix_HermiticityCondition} becomes $   \mathcal{K}\left(s\right)  \, = \,    \left[ \mathcal{K}\left( s\right)\right]^{*} $. 
  We also define fermionic derivatives to act from the left.
\bibliography{E_Jimenez_Super_Feyn_Rules}{}

\end{document}